\documentclass[aps,prc,preprint,showpacs,floatfix]{revtex4-1}
\usepackage{bm}
\usepackage{amsmath}
\usepackage{graphicx}
\usepackage{rotating}
\usepackage{subfigure}

\def\lazz{\mathrel{\mathchoice {\vcenter{\offinterlineskip\halign{\hfil
$\displaystyle##$\hfil\cr<\cr\sim\cr}}}
{\vcenter{\offinterlineskip\halign{\hfil$\textstyle##$\hfil\cr<\cr\sim\cr}}}
{\vcenter{\offinterlineskip\halign{
\hfil$\scriptstyle##$\hfil\cr<\cr\sim\cr}}}
{\vcenter{\offinterlineskip\halign{\hfil$\scriptscriptstyle##
$\hfil\cr<\cr\sim\cr}}}}}
\def\gazz{\mathrel{\mathchoice {\vcenter{\offinterlineskip\halign{\hfil
$\displaystyle##$\hfil\cr>\cr\sim\cr}}}
{\vcenter{\offinterlineskip\halign{\hfil$\textstyle##$\hfil\cr>\cr\sim\cr}}}
{\vcenter{\offinterlineskip\halign{
\hfil$\scriptstyle##$\hfil\cr>\cr\sim\cr}}}
{\vcenter{\offinterlineskip\halign{\hfil$\scriptscriptstyle##
$\hfil\cr>\cr\sim\cr}}}}}		
\def\pr{\prime}
\def\be{\begin{equation}}
\def\lan{\left\langle}
\def\ran{\right\rangle}
\def\ee{\end{equation}}
\def\barr{\begin{array}}
\def\earr{\end{array}}

\def\l{\left}
\def\r{\right}
\def\dis{\displaystyle}
\def\ed{\end{document}}

\def\bee{{\mbox{\boldmath $E$}}}

\def\we{{\widehat {E}}}
\def\wD{{\widetilde {D}}}
\def\wmp{{\widetilde {m}_1}}
\def\wmn{{\widetilde {m}_2}}

\def\cg{{\cal G}}

\def\bd{{\widetilde {\cal D}}}
\def\cd{{\cal D}}
\def\cx{{\cal X}}

\begin{document}

\title{One plus two-body random matrix ensembles with parity:  Density of
states and parity ratios}

\author{Manan Vyas$^1$, V.K.B. Kota$^{1,2,}$
\footnote{ Corresponding author, phone:
91-79-26314464, Fax: +91-79-26314460 \\ {\it E-mail address:} 
vkbkota@prl.res.in (V.K.B. Kota)} and P.C. Srivastava$^1$}

\affiliation{$^1$Physical Research Laboratory, Ahmedabad 380 009, India \\
$^2$Department of Physics, Laurentian University, Sudbury, Ontario, Canada
P3E 2C6}

\begin{abstract}

One plus two-body embedded Gaussian orthogonal ensemble of random matrices
with parity [EGOE(1+2)-$\pi$] generated by a random two-body interaction
(modeled by GOE in two particle spaces) in the presence of a mean-field, for
spinless identical  fermion systems, is defined, generalizing  the  two-body
ensemble with parity analyzed by Papenbrock and Weidenm\"{u}ller [Phys. Rev.
C  {\bf 78}, 054305 (2008)], in terms of two mixing parameters and a gap
between the positive $(\pi=+)$ and negative $(\pi=-)$ parity single particle
(sp) states. Numerical calculations are used to demonstrate, using realistic
values of the mixing parameters appropriate for some nuclei, that the
EGOE(1+2)-$\pi$ ensemble generates Gaussian form (with corrections) for
fixed parity eigenvalue  densities (i.e. state densities). The random matrix
model also generates many features in parity ratios of state densities that
are similar to those predicted by a method based on the Fermi-gas model for
nuclei. We have also obtained, by applying the formulation due to Chang et
al [Ann. Phys. (N.Y.) {\bf 66}, 137 (1971)], a simple formula for the
spectral variances defined over fixed-$(m_1,m_2)$ spaces, where $m_1$ is the
number of fermions in the $+$ve parity sp states and $m_2$ is the number of
fermions in the $-$ve parity sp states. Similarly, using the binary
correlation approximation, in the dilute limit, we have derived expressions
for the lowest two shape parameters.  The smoothed densities generated by
the sum of fixed-$(m_1,m_2)$ Gaussians with lowest two shape corrections
describe the numerical results in many situations.  The model  also
generates preponderance of $+$ve  parity ground states for small  values of
the mixing parameters and this is a feature seen in nuclear shell model
results. 

\end{abstract}

\pacs{21.60.Cs, 24.60.Lz, 21.10.Hw, 21.10.Ma}

\maketitle
\date{January 10, 2011}

\section{Introduction}

Random matrix theory (RMT), starting with Wigner and Dyson's Gaussian random
ensembles \cite{Po-65,Me-04} introduced to describe neutron resonance data
\cite{Br-81,Haq-82}, has emerged as a powerful statistical approach leading
to paradigmatic models describing generic properties of complex systems
\cite{Ze-96,Gu-98,Ul-08,Ha-10,Go-10}.  Developments and applications of RMT
in nuclear physics in last 30 years have been reviewed recently by
Weidenm\"{u}ller and  collaborators \cite{Mit-09,Mit-10}. The Wigner-Dyson
classical Gaussian orthogonal (GOE), unitary (GUE) and symplectic (GSE)
ensembles are ensembles of multi-body interactions while the nuclear
interparticle interactions are essentially two-body in nature. This together
with nuclear shell model examples led to the introduction of random matrix
ensembles generated by two-body interactions  in 1970-1971
\cite{Fr-70,Bo-71}. These two-body ensembles are defined by representing the
two-particle Hamiltonian by one  of the classical ensembles and then the $m$
($m>2$)  particle $H$-matrix is generated by the Hilbert space geometry.
Thus the random matrix ensemble in the two-particle spaces is embedded in
the $m$-particle $H$-matrix and therefore these ensembles are generically
called embedded ensembles (EEs). Simplest of these ensembles is the embedded
Gaussian orthogonal ensemble of random matrices generated by two-body
interactions for spinless fermion (boson) systems, denoted by EGOE(2)
[BEGOE(2); here `B' stands for bosons]. In addition to the complexity
generating two-body interaction, Hamiltonians for realistic systems such as
nuclei  consist of a mean-field one-body part.  Then the appropriate random
matrix ensembles are EE(1+2). The spinless fermion/boson EGEs (orthogonal
and unitary versions) have been explored in detail from 70's with a major
revival from 1995 and it is now well understood that EGEs model many-body
chaos or stochasticity exhibited by isolated finite interacting quantum
systems \cite{Ko-01,Go-10}.  Besides the mean-field and the two-body
character, realistic Hamiltonians also carry a variety of symmetries. In
many applications of EGEs, generic properties of EGEs for spinless fermions
are `assumed'  to extend to symmetry subspaces \cite{KH-10}.  More
importantly,  there are several  properties of real systems that require
explicit inclusion of symmetries and they are defined by a variety of Lie
algebras. This led to studies on EGEs with symmetries such as spin
\cite{JS-01,Pa-02,Ma-10,Ma-09,Ma-11}, spin-isospin $SU(4)$ \cite{Ma-10b},
$J$-symmetry \cite{Pa-07} and many others (see for example
\cite{Al-05,Ku-00}). In the present paper, we consider parity symmetry in EE
as there are several nuclear structure quantities that require explicit
inclusion of parity. Some of these are as follows.

Parity ratios of nuclear level densities is an important ingredient in
nuclear astrophysical applications. Recently, a method based on
non-interacting Fermi-gas model for proton-neutron systems has been
developed and the parity $(\pi)$ ratios as a function of excitation energy
in large number of nuclei of astrophysical interest have been tabulated
\cite{Mo-07}. The method is based on the assumption that the probability to
occupy $s$ out of $N$ given  single particle (sp) states follow Poisson
distribution in the dilute limit $(m << N, N\to \infty$ where $m$ is the
number of particles). Then the ratio of the partition functions for the
$+$ve and $-$ve parity states is given by the simple formula $Z_-/Z_+ =
\tanh f$, where $f$ is average number of particles in the $+$ve parity
states. Starting with this, an iterative method is developed with inputs
from the  Fermi-Dirac distribution for occupancies including pairing effects
and the Fermi-gas form for the total level density. In the examples studied
in \cite{Mo-07}, parity ratios are found to equilibrate only around $5-10$
MeV excitation energy. However, ab-initio interacting particle theory for
parity ratios is not yet available. 

A closely related question is about the form of the density of states
defined over spaces with fixed-$\pi$. In general, fixed-$\pi$ density of
states can be written as a sum of appropriate partial densities.  In the
situation that the form of the partial densities is determined by a few
parameters (as it is with a Gaussian or a Gaussian with one or two
corrections), it is possible to derive a theory for these parameters and
using these, one can construct fixed-$\pi$ density of states and calculate
parity ratios. Such a theory with interactions in general follows from
random matrix theory \cite{KH-10}. 

In addition to the questions related to fixed-$\pi$ density of states and
parity ratios, there is also the important recognition in the past few years
that random interactions generate regular  structures
\cite{Ze-04,Zh-04,Pa-07}. It was shown in \cite{Zh-04a} that shell model for
even-even nuclei gives  preponderance of  $+$ve parity ground states. A
parameter-free EGOE with parity has been  defined and analyzed recently
\cite{Pa-08} to address the question of `preponderance of ground states with
positive parity' for systems with even number of fermions. They show that in
the dilute limit, $+$ve parity ground states appear with only 50\%
probability. Thus, a random matrix theory describing shell model results is
not yet available.

With the success of the embedded random matrix ensembles (EE)
\cite{Ko-01,Go-10}, one can argue that the EE generated by parity preserving
random interaction  may provide generic results for the three nuclear
structure quantities mentioned above. For nuclei, the GOE versions of EE are
relevant. Then, with a random (modeled by GOE) two-body interaction
preserving parity in the presence of a mean-field, we have embedded Gaussian
orthogonal ensemble of one plus two-body interactions with parity [hereafter
called  EGOE(1+2)-$\pi$]. This model contains two mixing parameters and a
gap between the  $+$ve and $-$ve parity sp states and it goes much beyond
the simpler model considered in \cite{Pa-08}. In the random matrix model
used in the present paper, proton-neutron degrees of freedom and  angular
momentum $(J)$ are not considered. Let us add that in the present paper for
the first time a random matrix theory for parity ratios is attempted.  Now
we will give a preview. 

Section II gives the definition of EGOE(1+2)-$\pi$ and a method for its
construction. From the results known for EE for spinless fermion (boson)
systems, for fermions (bosons) with spin and from shell model calculations
\cite{Br-81,Ko-01,Ma-10}, it is expected that the fixed-$\pi$ state
densities (more appropriately partial densities) approach Gaussian form in
general. Therefore, exact propagation formulas for fixed-$\pi$ energy
centroids and spectral variances are derived and the results are given in
Section III. Used here is the group theoretical formulation developed by
Chang et al \cite{CFT}. Similarly in Appendix B,  given are the formulas for
the ensemble averaged skewness $\gamma_1(m_1,m_2)$ and excess
$\gamma_2(m_1,m_2)$ parameters for fixed-$(m_1,m_2)$ partial densities (with
$m_1$ fermions distributed in $N_+$ number of $+$ve parity sp levels and
similarly $m_2$ fermions in $N_-$ number of $-$ve parity sp levels) and used
is the binary correlation approximation method described in
\cite{Mo-73,Mo-75,To-86,Fr-88}. These will provide corrections to the
Gaussian state densities. In Section IV, presented are the numerical results
for (i) fixed-$\pi$ state densities, (ii) parity ratios of state densities
and (iii) probability for $+$ve parity ground states. Finally, Section V
gives conclusions and future outlook. 

\section{EGOE(1+2)-$\pi$ ensemble}

\begin{figure}
\includegraphics[width=4in,height=2.5in]{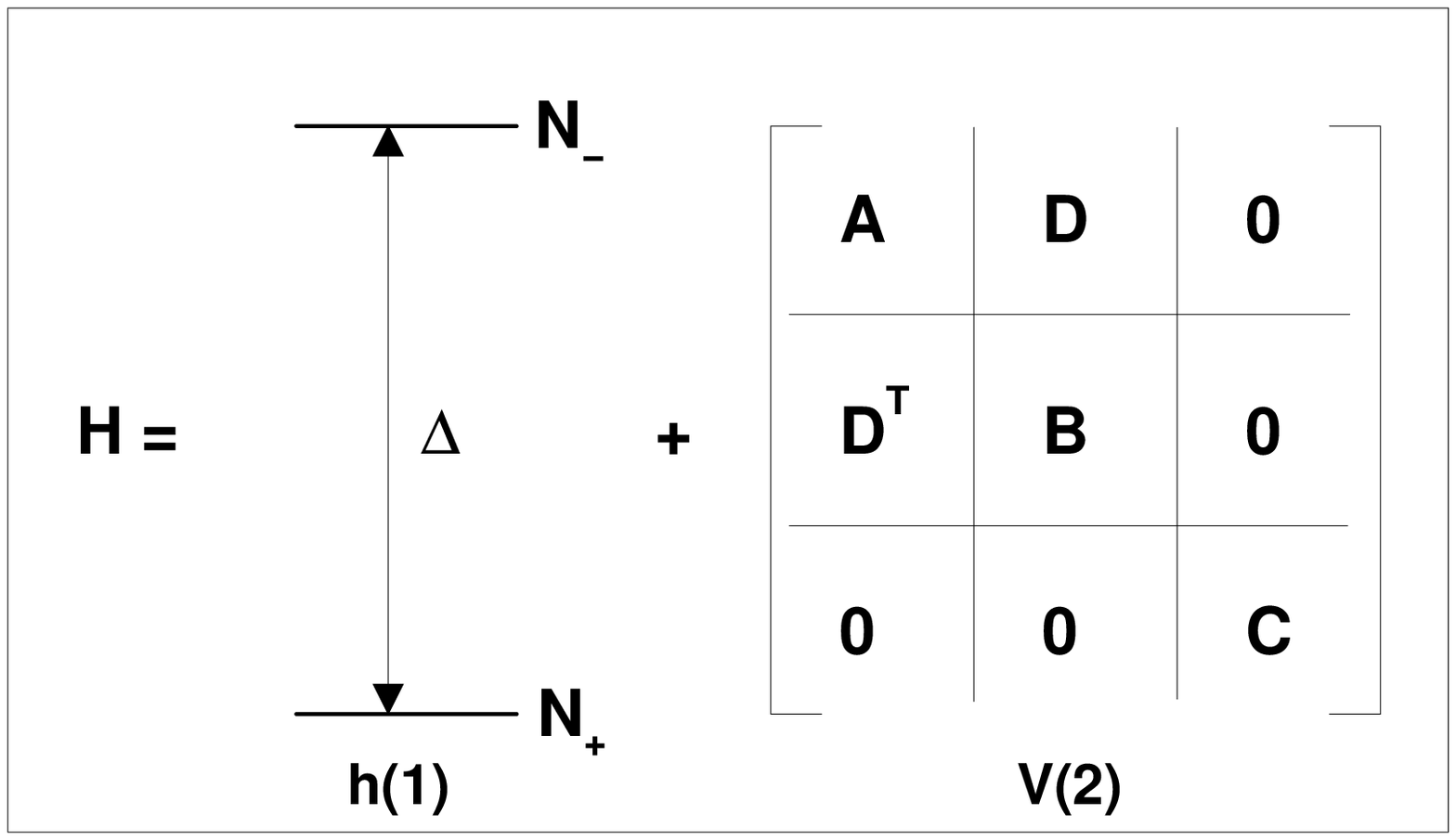}
\caption{Parity preserving one plus two-body $H$ with a sp spectrum 
defining $h(1)$ along with a schematic form of the $V(2)$ matrix. 
Dimension of the matrices A, B and C are $N_+(N_+-1)/2$, $N_-(N_--1)/2$ 
and $N_+N_-$ respectively. Note that $D^T$ is the transpose of the matrix 
$D$. See text for details.}
\label{fig1}
\end{figure}

Given $N_+$ number of positive parity sp states and  similarly $N_-$ number
of negative parity sp states, let us assume, for simplicity, that the $+$ve
and $-$ve parity states are degenerate and separated by energy $\Delta$ (see
Fig. \ref{fig1}). This defines the one-body part $h(1)$ of the Hamiltonian
$H$ with $N=N_+ + N_-$ sp states. The matrix for the two-body part $V(2)$ of
$H$ [we assume $H$ is (1+2)-body] will be a $3 \times 3$ block matrix in two
particle spaces as there are three possible ways to generate two particle
states with definite parity:  (i) both fermions  in $+$ve parity states;
(ii) both fermions in $-$ve parity states; (iii) one fermion  in $+$ve and
other fermion in $-$ve parity states. They will give the matrices $A$, $B$
and $C$ respectively in Fig. \ref{fig1}. For parity  preserving interactions
only the states (i) and (ii) will be mixed and mixing matrix is $D$ in Fig.
\ref{fig1}. Note that the matrices $A$, $B$ and $C$ are symmetric square
matrices while $D$ is in  general a rectangular mixing matrix. Consider $N$
sp states arranged such that the states $1$ to $N_+$ have $+$ve parity and 
states $N_++1$ to $N$ have $-$ve parity. Then the operator form of $H$ 
preserving parity is,
\be
\barr{rcl}
H & = & h(1) + V(2)\;; \\ \\
h(1) & = & \dis\sum_{i=1}^{N_+} \epsilon_i^{(+)} \hat{n}_i^{(+)} + 
\dis\sum_{i=N_++1}^{N} \epsilon_i^{(-)} \hat{n}_i^{(-)} \;;\;\;\;\; 
\epsilon_i^{(+)}=0 \;,\;\; \epsilon_i^{(-)}=\Delta\;, \\ \\
V(2) & = & 
\dis\sum_{\barr{c}
i,j,k,l=1 \\
(i<j,\; k<\ell) \earr}^{N_+} 
\lan \nu_k \; \nu_\ell \mid V \mid \nu_i \; \nu_j
\ran a^\dagger_k \; a^\dagger_\ell \; a_j \; a_i \\ \\
& + & 
\dis\sum_{\barr{c} 
i^\pr,j^\pr,k^\pr,\ell^\pr=N_++1 \\
(i^\pr<j^\pr,\; k^\pr<\ell^\pr)
\earr}^N 
\lan \nu_{k^\pr} \; \nu_{\ell^\pr} \mid V \mid \nu_{i^\pr} \; \nu_{j^\pr}
\ran a^\dagger_{k^\pr} \; a^\dagger_{\ell^\pr} \; a_{j^\pr} \; a_{i^\pr}
\\ \\
& + & 
\dis\sum_{
i^{\pr\pr},k^{\pr\pr}=1}^{N_+}\;\; 
\dis\sum_{j^{\pr\pr},\ell^{\pr\pr}=N_++1}^N
\lan \nu_{k^{\pr\pr}} \; \nu_{\ell^{\pr\pr}} \mid V \mid \nu_{i^{\pr\pr}} 
\; \nu_{j^{\pr\pr}} \ran a^\dagger_{k^{\pr\pr}} \; a^\dagger_{\ell^{\pr\pr}}
 \; a_{j^{\pr\pr}} \; a_{i^{\pr\pr}} \\ \\
& + &
\dis\sum_{\barr{c} P,Q=1 \\ (P<Q) \earr}^{N_+}\;\;
\dis\sum_{\barr{c} R,S=N_++1 \\ (R<S) \earr}^N
\l[ \lan \nu_P \; \nu_Q \mid V \mid \nu_R \; \nu_S \ran a^\dagger_P \;
a^\dagger_Q \; a_S \; a_R + \mbox{h.c.} \r]\;.
\earr \label{eq.ham}
\ee
In Eq. (\ref{eq.ham}), $\nu_i$'s are sp states with $i=1,2,\ldots,N$ (the
first $N_+$ states are $+$ve parity and remaining $-$ve parity). Similarly,
$\lan \ldots \mid V \mid \ldots \ran$ are the two-particle matrix elements,
$\hat{n}_i$ are number operators and $a^\dagger_i$ and $a_i$ are creation
and annihilation operators respectively. Note that the four terms in the RHS
of the expression for $V(2)$ in  Eq.  (\ref{eq.ham}) correspond respectively
to the matrices $A$, $B$, $C$ and $D$ shown in Fig. \ref{fig1}.

Many particle states for $m$ fermions in the $N$ sp states can be obtained
by distributing $m_1$ fermions in the $+$ve parity sp states ($N_+$ in
number)  and similarly, $m_2$ fermions in the $-$ve parity sp states ($N_-$
in number) with $m=m_1+m_2$. Let us denote each distribution of $m_1$
fermions in $N_+$ sp states by $\bf{m}_1$ and similarly, $\bf{m}_2$ for
$m_2$ fermions in  $N_-$ sp states. Many particle basis defined by
$(\bf{m}_1, \bf{m}_2)$ with $m_2$ even will form the basis for $+$ve parity
states and similarly, with $m_2$ odd  for $-$ve parity states. In the
$(\bf{m}_1, \bf{m}_2)$ basis with $m_2$ even (or odd), the $H$ matrix
construction reduces to the matrix construction for spinless  fermion
systems. The method of construction for spinless fermion systems is well
known \cite{Ko-01} and therefore it is easy to construct the many particle 
$H$ matrices in $+$ve and $-$ve parity spaces. The matrix dimensions $d_+$
for $+$ve parity and $d_-$ for $-$ve parity spaces are given by,
\be
d_+ = \dis\sum_{m_1,m_2\;(m_2\;even)} 
\dis\binom{N_+}{m_1} \dis\binom{N_-}{m_2}
\;,\;\;\;\; 
d_- = \dis\sum_{m_1,m_2\;(m_2\;odd)} 
\dis\binom{N_+}{m_1} \dis\binom{N_-}{m_2}
\;.
\label{eq.dim}
\ee
Some examples for the dimensions $d_+$ and $d_-$ are given in Table
\ref{dim}.

\begin{table}[htp]
\caption{Hamiltonian matrix dimensions $d_+$ and $d_-$ for various 
values of $(N_+,N_-,m)$.}
\begin{center}
\resizebox { 0.7\textwidth }{ 0.3\textwidth }{
\begin{tabular}{|c|c|c|c|c||c|c|c|c|c|}
\hline
$N_+$ & $N_-$ & $m$ & $d_+$ & $d_-$ & $N_+$ & $N_-$ & $m$ & $d_+$ & $d_-$ 
\\ 
\hline
$6$ & $6$ & $6$ & $452$ & $472$ & $8$ & $8$ & $4$ & $924$ & $896$ \\
$7$ & $5$ & $6$ & $462$ & $462$ & $$ & $$ & $5$ & $2184$ & $2184$ \\
$7$ & $7$ & $5$ & $1001$ & $1001$ & $$ & $$ & $6$ & $3976$ & $4032$ \\
$$ & $$ & $6$ & $1484$ & $1519$ & $10$ & $6$ & $4$ & $900$ & $920$ \\
$$ & $$ & $7$ & $1716$ & $1716$ & $$ & $$ & $5$ & $2202$ & $2166$ \\
$8$ & $6$ & $5$ & $1016$ & $986$ & $$ & $$ & $6$ & $4036$ & $3972$ \\
$$ & $$ & $6$ & $1499$ & $1504$ & $6$ & $10$ & $4$ & $900$ & $920$ \\
$9$ & $5$ & $5$ & $1011$ & $911$ & $$ & $$ & $5$ & $2166$ & $2202$ \\
$$ & $$ & $6$ & $1524$ & $1479$ & $$ & $$ & $6$ & $4036$ & $3972$ \\
$5$ & $10$ & $4$ & $665$ & $700$ & $9$ & $9$ & $6$ & $9240$ & $9324$ \\
$$ & $$ & $5$ & $1501$ & $1502$ & $10$ & $8$ & $6$ & $9268$ & $9296$ \\
$$ & $$ & $$ & $$ & $$ & $10$ & $10$ & $5$ & $7752$ & $7752$ \\
$$ & $$ & $$ & $$ & $$ & $$ & $$ & $6$ & $19320$ & $19440$ \\
\hline
\end{tabular}}
\end{center}
\label{dim}
\end{table}

The EGOE(1+2)-$\pi$ ensemble is defined by choosing the matrices $A$, $B$
and  $C$ to be independent GOE's with matrix elements variances $v_a^2$,
$v_b^2$ and $v_c^2$ respectively. Similarly the matrix elements of the
mixing $D$ matrix are chosen to be independent (independent of $A$, $B$ and
$C$ matrix elements) zero centered  Gaussian variables with variance
$v_d^2$. Without loss of generality we choose $\Delta=1$ so that all the
$v$'s are in $\Delta$ units. This general EGOE(1+2)-$\pi$ model will have
too many parameters ($v_a^2,v_b^2,v_c^2,v_d^2,N_+,N_-,m$) and therefore it
is necessary to reduce the number of parameters. A numerically tractable and
physically relevant (as discussed ahead) restriction is to choose the  
matrix elements variances of the diagonal blocks $A$, $B$ and $C$ to be same
and then  we have the EGOE(1+2)-$\pi$ model defined by ($N_+,N_-,m$) and the
variance parameters ($\tau$,$\alpha$) where
\be
\dis\frac{v_a^2}{\Delta^2} = \dis\frac{v_b^2}{\Delta^2} =
\dis\frac{v_c^2}{\Delta^2} = \tau^2 \;,\;\;\;\;
\dis\frac{v_d^2}{\Delta^2} = \alpha^2 \;. 
\label{eq.taual}
\ee
Thus EGOE(1+2)-$\pi$ we employ is 
\be
\barr{l}
A:\; \mbox{GOE}(0:\tau^2)\;, \;B:\; \mbox{GOE}(0:\tau^2)\;, \;C:\; 
\mbox{GOE}(0:\tau^2)\;,\;D:\;\mbox{GOE}(0:\alpha^2)\;;\\
A,\;B,\;C,\;D \mbox{\;are\;independent\;GOE's}\;.
\earr\label{eq.model}
\ee
Note that the $D$ matrix is a GOE only in the sense that the matrix elements
$D_{ij}$ are all independent zero centered Gaussian variables with variance
$\alpha^2$. In the limit $\tau^2 \rightarrow \infty$ and $\alpha = \tau$,
the model defined by Eqs. (\ref{eq.ham}), (\ref{eq.taual}) and
(\ref{eq.model})  reduces to the simpler model analyzed in \cite{Pa-08}.

Before proceeding further, it is useful to mention that we are considering
in this paper spinless fermion systems (with parity included) just as in the
previous investigation \cite{Pa-08}. It is possible to extend the ensemble
to nucleons in shell model $j$-orbits (including both $+$ve and $-$ve orbits
such as for example $sd$ and $fp$) and construct the ensemble in
many-nucleon spaces with a given $J^\pi$ or $J^{\pi}T$ using  a shell model
code. However, such an attempt has not been made, just as in \cite{Pa-08},
as our focus is on parity. Also the ensemble for spinless systems will give
the essential features due to parity and these can be  used in later
explorations using ensembles with $J^\pi$ or $J^{\pi}T$ which are more
complicated numerically and more importantly from analytical point of view 
\cite{Br-81,Pa-07}. In fact, due to the severe problems associated with
analytical tractability, a variety of EGOE are being analyzed since 1995;
see \cite{Ko-01,Go-10,Pa-07} for reviews.  At this point it is also useful
to mention that EGOE(1+2)'s are also called TBRE in literature; see Section
5.7 in \cite{Go-10} for clarifications on this nomenclature. As Brody et al
state \cite{Br-81}: {\it The most severe mathematical difficulties with TBRE
are due to angular momentum constraints $\ldots$ Another type of ensemble,
$\ldots$ much closer to being mathematical tractable abandons the $J$
restrictions entirely $\ldots$ an embedded GOE, or EGOE for short.}   

Starting with the EGOE(1+2)-$\pi$ ensemble defined by Eqs. (\ref{eq.ham}),
(\ref{eq.taual}) and (\ref{eq.model}), we have numerically constructed 100
(in some examples 200) members of the ensemble in many-particle $+$ve and 
$-$ve parity spaces with dimensions $d_+$ and $d_-$ given by Eq.
(\ref{eq.dim}) for several values of ($N_+,N_-,m$) and varying the
parameters $\tau$ and $\alpha$.  This means we have considered 100
realizations of EGOE(1+2)-$\pi$ random matrices in ($N_+,N_-,m$) spaces - we
use the phrase `members' throughout the paper instead of `realizations'
(other names used by some authors are  `sets', `samples' and `trials') as in
all our previous papers. Before discussing the numerical calculations, we
present the results for the energy centroids, variances and also the shape
parameters (skewness and excess) defining the normalized fixed-$(m_1,m_2)$ 
partial densities
\be
\barr{l}
\rho^{m_1,m_2}(E) = \lan \delta (H - E) \ran^{m_1,m_2} =
\dis\frac{1}{d(m_1,m_2)} \dis\sum_{\alpha,\beta} \l|
C^{m_1,m_2,\alpha}_{E,\beta} \r|^2\;; \\ \\
\l| m_1,m_2,\alpha \ran = 
\dis\sum_\beta C^{m_1,m_2,\alpha}_{E,\beta} \l| E,\beta 
\ran \;,
\earr \label{eq.denpty1}
\ee
where, $\lan \ldots \ran$ corresponds to average and $\alpha$ and  $\beta$
are extra labels required to specify completely the states with a given
$(m_1,m_2)$ and $E$ respectively. Later we use the symbol $\lan \lan \ldots
\ran \ran$ that denotes the trace. These will allow us to understand some of
the numerical results. Let us add that the fixed-$\pi$ eigenvalue densities
$I_\pm(E)$ are sum of the appropriate partial densities as given by Eq.
(\ref{eq.densty}) ahead. Note that the densities $I_\pm(E)$ are normalized
to $d_\pm$.

\section{Energy centroids, variances, skewness and excess parameters 
for fixed-$(m_1,m_2)$ partial densities}

Let us call the set of $+$ve parity sp states as  unitary orbit \#1 and
similarly the set of $-$ve parity sp states as  unitary orbit \#2; see
\cite{KH-10} for unitary orbits notation and significance. For convenience,
from now on, we denote the sp states by the roman letters $(i,j,\ldots)$ and
unitary orbits by greek letters $(\alpha,\beta,\ldots)$. Note that
$\alpha=1$ corresponds to the $+$ve parity unitary orbit and $\alpha=2$
corresponds to the $-$ve parity unitary orbit (with this notation, $N_1=N_+$
and $N_2=N_-$). The sp states that belong to a unitary orbit $\alpha$ are
denoted as $i_\alpha,j_\alpha,\ldots$. Propagation formulas for the energy
centroids and variances of the partial densities $\rho^{m_1,m_2}(E)$ follow
from the unitary decomposition of $V(2)$ with respect to the sub-algebra
$U(N_+) \oplus U(N_-)$ contained in $U(N)$. Note that $(m_1,m_2)$ label the 
irreducible representations (irreps) of  $U(N_+) \oplus U(N_-)$ and they all
belong to the $U(N)$ irreps labeled by $m$. The $(m_1,m_2)$ are often called
unitary configurations \cite{KH-10}. With respect to $U(N_+) \oplus U(N_-)$,
the operator $V(2)$ decomposes into three parts $V(2) \to V^{[0]} + V^{[1]}
+V^{[2]}$. The $V^{[0]}$ generates the energy centroids $\lan V
\ran^{m_1,m_2}$, $V^{[1]}$ corresponds to the `algebraic' mean-field
generated by $V$ and  $V^{[2]}$ is the remaining irreducible two-body part.
Extending the  unitary decomposition for the situation with a single orbit
for spinless fermions (given in Appendix A) and also using the detailed
results in \cite{CFT}, we obtain the following formulas for the
$V^{[\nu]}$'s. The $V^{[0]}$ is given by (with $\alpha = 1,\;2$ and $\beta =
1,\;2$)
\be
\barr{rcl}
V^{[0]} & = & 
\dis\sum_{\alpha \geq \beta} 
\dis\frac{\hat{n}_\alpha (\hat{n}_\beta -
\delta_{\alpha\beta})}{(1+\delta_{\alpha\beta})}\; V_{\alpha\beta}\;;
\\ \\
V_{\alpha\alpha} & = &
\dis\binom{N_\alpha}{2}^{-1}\;\dis\sum_{i>j}V_{i_\alpha j_\alpha i_\alpha
j_\alpha}\;,\\ \\
V_{\alpha\beta} & = &
\l(N_\alpha N_\beta\r)^{-1}\;\dis\sum_{i,j}V_{i_\alpha j_\beta i_\alpha
j_\beta}\;;\;\;\;\;\alpha \neq \beta\;.
\earr \label{eq.nu0}
\ee
Then the traceless part
$\widetilde{V} = V -V^{[0]} = V^{[1]} + V^{[2]}$ where 
$(\widetilde{V})_{i_\alpha j_\beta i_\alpha j_\beta} = 
V_{i_\alpha j_\beta i_\alpha j_\beta} - V_{\alpha\beta}$ and 
$(\widetilde{V})_{ijk\ell} = V_{ijk\ell}$ for all others.
Now the $V^{[1]}$ part is
\be
\barr{l}
V^{[1]} = \dis\sum_{\alpha,i,j} 
\widehat{\xi}_{i_\alpha j_\alpha}
a^\dagger_{i_\alpha}\;a_{j_\alpha}\;;\\ \\
\widehat{\xi}_{i_\alpha j_\alpha} = 
\dis\sum_{\beta} \dis\frac{\hat{n}_\beta -
\delta_{\alpha\beta}}{N_\beta - 2\delta_{\alpha\beta}} \; \zeta_{i_\alpha
j_\alpha} (\beta)\;,\;\;\;\; 
\zeta_{i_\alpha j_\alpha} (\beta) = 
\dis\sum_{k_\beta} \widetilde{V}_{k_\beta i_\alpha k_\beta
j_\alpha}\;.
\earr \label{eq.nu1}
\ee
Finally, the $V^{[2]}$ part is as follows,
\be
\barr{rcl}
V^{[2]} & = & \widetilde{V} - V^{[1]}\;; \\ \\
{V}^{[2]}_{i_\alpha j_\beta i_\alpha j_\beta} & = &
\widetilde{V}_{i_\alpha j_\beta i_\alpha j_\beta} - \l[ 
\dis\frac{\zeta_{i_\alpha j_\alpha} (\beta)}
{N_\beta - 2\delta_{\alpha\beta}}
+ \dis\frac{\zeta_{i_\beta j_\beta} (\alpha)}
{N_\alpha - 2\delta_{\alpha\beta}}
 \r] \;,\\ \\
{V}^{[2]}_{k_\alpha i_\beta k_\alpha j_\beta} & = &
\widetilde{V}_{k_\alpha i_\beta k_\alpha j_\beta} - 
\dis\frac{\zeta_{i_\beta j_\beta} (\alpha)}
{N_\alpha - 2\delta_{\alpha\beta}}
\;;\;\;\;\; i_\beta \neq j_\beta \;,\\ \\
{V}^{[2]}_{i j k \ell} & = &
\widetilde{V}_{i j k \ell} \mbox{\;\;for\;all\;others}\;.
\earr \label{eq.nu2}
\ee
Given Eqs. (\ref{eq.nu0}), (\ref{eq.nu1}) and (\ref{eq.nu2}),  by intuition
and  using Eq. (\ref{eq.a3}), it is possible to write the propagation
formulas for the energy centroids and variances of $\rho^{m_1,m_2}(E)$. Note
that these are essentially traces of $H$ and $H^2$ over the space defined by
the two-orbit configurations $(m_1,m_2)$; see Eqs.  (\ref{eq.cent}) and
(\ref{eq.var}) ahead. A direct approach to write the propagation formulas
for centroids and variances for a multi-orbit configuration was given in
detail  first by French and Ratcliff \cite{Fr-71}. The formula for the
variance given in \cite{Fr-71} is cumbersome  and it is realized later
\cite{CFT} that they can be made compact by applying group theory (see also
\cite{Ko-01,Wo-86,KH-10}).   We have adopted the group theoretical approach
for the two-orbit averages and obtained formulas. Propagation formula for
the fixed-$(m_1,m_2)$ energy centroids is,
\be
E_c(m_1,m_2) = \lan H \ran^{m_1,m_2} = m_2\;\Delta + 
\dis\sum_{\alpha \geq \beta} 
\dis\frac{m_\alpha (m_\beta -
\delta_{\alpha\beta})}{(1+\delta_{\alpha\beta})}\; V_{\alpha\beta} \;.
\label{eq.cent}
\ee
First term in Eq. (\ref{eq.cent})  is generated by $h(1)$ and it is simple
because of the choice of the sp energies as shown in Fig. \ref{fig1}.
Propagation formula for fixed-$(m_1,m_2)$ variances is,
\be
\barr{rcl}
\sigma^2(m_1,m_2) & = &  
\lan H^2 \ran^{m_1,m_2} - \l[ \lan H \ran^{m_1,m_2}\r]^2 \\ \\ 
& = & 
\dis\sum_\alpha \dis\frac{m_\alpha\l( N_\alpha -
m_\alpha \r)}{N_\alpha \l( N_\alpha -1 \r)} \dis\sum_{i_\alpha, j_\alpha}
\l[{\xi}_{i_\alpha, j_\alpha}(m_1,m_2)\r]^2 \\ \\
& + & \dis\sum^\pr_{\alpha,\beta,\gamma,\delta} 
\dis\frac{m_\alpha(m_\beta-\delta_{\alpha\beta})(N_\gamma -
m_\gamma)(N_\delta-m_\delta-\delta_{\gamma\delta})}
{N_\alpha(N_\beta - \delta_{\alpha\beta})
(N_\gamma - \delta_{\gamma\alpha} - \delta_{\gamma\beta})
(N_\delta - \delta_{\delta\alpha} - \delta_{\delta\beta} 
- \delta_{\delta\gamma})} \; (X) \;;\\ \\
{\xi}_{i_\alpha, j_\alpha}(m_1,m_2) & = & 
\dis\sum_{\beta} \dis\frac{m_\beta -
\delta_{\alpha\beta}}{N_\beta - 2\delta_{\alpha\beta}} \; \zeta_{i_\alpha
j_\alpha} (\beta)\;, \;\;\;\;
X = \dis\sum^\pr \l( V^{[2]}_{i_\alpha
j_\beta k_\gamma \ell_\delta} \r)^2 \;.
\earr \label{eq.var}
\ee
The `prime' over summations in Eq. (\ref{eq.var}) implies that the
summations are not free sums. Note that $(\alpha,\beta,\gamma,\delta)$ take
values $(1,1,1,1)$, $(2,2,2,2)$, $(1,2,1,2)$, $(1,1,2,2)$ and $(2,2,1,1)$.
Similarly, in the sum over $(i_\alpha,j_\beta)$, $i \leq j$ if
$\alpha=\beta$ and otherwise the sum is over all $i$ and $j$. Similarly,
for  $(k_\gamma, \ell_\delta)$. Using $E_c(m_1,m_2)$ and
$\sigma^2(m_1,m_2)$,  the  fixed-parity energy centroids and spectral 
variances [they define $I_\pm(E)$] can be obtained as follows,
\be
\barr{l}
E_c(m,\pm) = \lan H \ran^{m,\pm} = \dis\frac{1}{d_\pm} \dis\sum_{
m_1,m_2}^\pr d(m_1,m_2) E_c(m_1,m_2)\;, \\ \\
\sigma^2(m,\pm) = \lan H^2 \ran^{m,\pm} - \l[ \lan H \ran^{m,\pm}\r]^2 
\;;\\ \\
\lan H^2 \ran^{m,\pm} = \dis\frac{1}{d_\pm} \dis\sum_{m_1,m_2}^\pr
d(m_1,m_2) \l[ \sigma^2(m_1,m_2) + E_c^2(m_1,m_2) \r]\;. 
\earr \label{eq.cenvar}
\ee
The `prime' over summations in Eq. (\ref{eq.cenvar}) implies that
$m_2$ is even(odd) for $+$ve($-$ve) parity.

It should be pointed out that the  formulas given by Eqs. (\ref{eq.cent}),
(\ref{eq.var}) and  (\ref{eq.cenvar}) are compact and easy to understand
compared to Eqs. (10)-(14) of \cite{Pa-08} and also those that follow from
Eqs. (129) and (133) of \cite{Fr-71} where unitary  decomposition is not
employed. We have verified Eqs. (\ref{eq.cent}) and (\ref{eq.var})  by
explicit construction of the $H$ matrices in many examples.  In principle,
it is possible to obtain a formula for the ensemble averaged variances
using  Eq. (\ref{eq.var}); the  ensemble averaged centroids derive only from
$h(1)$. Simple asymptotic formulas for ensemble  averaged variances follow
by neglecting the $\delta$-functions that appear in Eq. (\ref{eq.var}) and
replacing $\widetilde{V}^2_{i j k \ell}$ by $\tau^2$ and $\alpha^2$
appropriately. Then the final formula for the ensemble averaged
fixed-$(m_1,m_2)$ variances is,
\be
\barr{rcl}
\overline{\sigma^2(m_1,m_2)} & \approx & m \l[ \dis\sum_{\alpha=1}^2 
m_\alpha \l( N_\alpha - m_\alpha\r)  \r] \; \tau^2 \\ \\
& + & \l[ \dis\binom{m_1}{2} \dis\binom{\wmp}{2} + 
\dis\binom{m_2}{2} \dis\binom{\wmn}{2} +
m_1 m_2 \wmp \wmn \r] \;\tau^2 \\ \\
& + & \l[ \dis\binom{m_1}{2} \dis\binom{\wmn}{2} + 
\dis\binom{m_2}{2} \dis\binom{\wmp}{2} \r]\;\alpha^2\;.
\earr \label{eq.eavar}
\ee
Here, $\wmp = N_1 - m_1$ and $\wmn = N_2 - m_2$. The `overline' in Eq.
(\ref{eq.eavar}) denotes ensemble average. In Table \ref{widths}, we
compare the results obtained from Eq. (\ref{eq.eavar}) with those obtained
for various  100 member ensembles using Eq. (\ref{eq.var}) and the
agreements are quite good. Therefore, in many practical applications, one
can use Eq. (\ref{eq.eavar}). 

\begin{table}[htp]
\caption{Ensemble averaged  fixed-$(m_1,m_2)$ widths $\sigma(m_1,m_2)$  and
the total spectral width $\sigma_t$ for different $(\tau,\alpha)$ values.
For each $(\tau,\alpha)$, the $\sigma(m_1,m_2)$ are given  in the table and
they are obtained using the exact propagation formula Eq. (\ref{eq.var}) for
each member of the  ensemble. In all the calculations, ensembles with 100
members are employed.  Numbers in the bracket are obtained by using the
asymptotic formula given in Eq. (\ref{eq.eavar}). Last row for each
$(N_+,N_-)$ gives the corresponding $\sigma_t$  values. All the results are
given for 6 particle systems and the dimensions  $d(m_1,m_2)$ are also given
in the table. See text for details.}
\begin{center}
\resizebox { 0.8\textwidth }{ 0.45\textwidth }{
\begin{tabular}{|c|c|c|c|c|c|c|c|}
\hline
 & & & & \multicolumn{4}{c|}{$(\tau,\alpha/\tau)$} \\ 
\cline{5-8}
$(N_+,N_-)$ & $m_1$ & $m_2$ & $d(m_1,m_2)$ &
$(0.1,0.5)$ & $(0.1,1.5)$ & $(0.2,0.5)$ & $(0.2,1.5)$ \\ 
\hline
$(8,8)$ &  0 & 6  &    28  & 1.36(1.39) & 3.21(3.21)  & 2.73(2.77) 
& 6.41(6.42)  \\
        &  1 & 5  &   448  & 1.76(1.79) & 2.70(2.72)  & 3.52(3.57) 
	& 5.41(5.44)  \\
	&  2 & 4  &  1960  & 2.05(2.09) & 2.48(2.50)  & 4.11(4.17) 
	& 4.96(5.01)  \\
	&  3 & 3  &  3136  & 2.16(2.19) & 2.42(2.45)  & 4.31(4.38) 
	& 4.84(4.90)  \\
	&  4 & 2  &  1960  & 2.05(2.09) & 2.48(2.50)  & 4.11(4.17) 
	& 4.95(5.01)  \\
	&  5 & 1  &   448  & 1.76(1.79) & 2.70(2.72)  & 3.52(3.57) 
	& 5.41(5.44)  \\
	&  6 & 0  &    28  & 1.37(1.39) & 3.21(3.21)  & 2.75(2.77) 
	& 6.42(6.42)  \\
	&    &    &        & 2.29(2.32) & 2.68(2.71)  & 4.24(4.30) 
	& 5.08(5.13)  \\
\hline 
$(6,10)$ & 0 & 6  &  210   & 1.67(1.70) & 2.70(2.72) & 3.34(3.41) 
        & 5.41(5.44) \\
        &  1 & 5  &  1512  & 2.04(2.07) & 2.48(2.51) & 4.08(4.15) 
	& 4.97(5.02) \\
	&  2 & 4  &  3150  & 2.19(2.22) & 2.41(2.44) & 4.37(4.44) 
	& 4.82(4.88) \\
	&  3 & 3  &  2400  & 2.11(2.14) & 2.43(2.46) & 4.22(4.28) 
	& 4.86(4.91) \\
	&  4 & 2  &  675   & 1.84(1.87) & 2.60(2.62) & 3.67(3.73) 
	& 5.20(5.24) \\
	&  5 & 1  &  60    & 1.46(1.48) & 3.06(3.06) & 2.92(2.96) 
	& 6.12(6.13) \\
	&  6 & 0  &  1	   & 1.30(1.30) & 3.90(3.90) & 2.60(2.60) 
	& 7.81(7.79) \\
	&    &    &        & 2.31(2.33) & 2.65(2.67) & 4.30(4.36) 
	& 5.02(5.07) \\
\hline 
$(10,10)$ & 0 & 6  &   210  & 1.97(2.01) & 4.16(4.19) & 3.95(4.01) 
         & 8.33(8.37) \\
         &  1 & 5  &  2520  & 2.44(2.49) & 3.63(3.66) & 4.90(4.98) 
	 & 7.25(7.32) \\
	 &  2 & 4  &  9450  & 2.76(2.81) & 3.36(3.40) & 5.53(5.61) 
	 & 6.71(6.79) \\
	 &  3 & 3  & 14400  & 2.87(2.92) & 3.28(3.32) & 5.74(5.83) 
	 & 6.56(6.64) \\
	 &  4 & 2  &  9450  & 2.76(2.81) & 3.36(3.40) & 5.53(5.61) 
	 & 6.71(6.79) \\
	 &  5 & 1  &  2520  & 2.44(2.49) & 3.63(3.66) & 4.90(4.98) 
	 & 7.25(7.32) \\
	 &  6 & 0  &   210  & 1.97(2.01) & 4.16(4.19) & 3.95(4.01) 
	 & 8.33(8.37) \\
	 &    &    &        & 2.95(2.99) & 3.54(3.57) & 5.62(5.70) 
	 & 6.83(6.91) \\
\hline
\end{tabular} }
\end{center}
\label{widths}
\end{table}

The skewness and excess parameters $\gamma_1$ and $\gamma_2$ give
information about the shape of the partial densities and they are close to 
zero implies Gaussian form. Formulas for the $M_r$, $r=3,4$ for a given one
plus two-body  Hamiltonian defined by Eq. (1) follow from the results given
in \cite{Wo-86,No-72,Ay-74,Po-75,Ch-78,Ka-95} many years back. However,
these formulas contain very large  number of complicated terms (in
particular for $M_4$) and carrying out analytically ensemble averaging is
proved to be impractical (we are not aware if anyone was successful in the
past). Some idea of the difficulty in carrying out simplifications can be
seen from the attempt in \cite{Pl-97}.  An alternative is to program the
exact formulas and evaluate the moments numerically for  each member of
EGOE(1+2)-$\pi$ by considering say 500 members in two particle spaces. As
pointed out  by Ter\'{a}n and Johnson \cite{Jo-06a} in their most recent 
attempt, these calculations for the 4th moments are time consuming if not
impractical.  All the problems with the exact formulas have been emphasized
in \cite{KH-10}. Because of these (in future with much faster computers it
may be possible to use the exact formulas), we have adopted the binary
correlation approximation, first used by Mon and French \cite{Mo-73,Mo-75}
and later by French et al \cite{Fr-88,To-86},  which is good in the dilute
limit: $m_1, N_1, m_2, N_2 \to \infty$, $m/N_1 \to 0$ and $m/N_2 \to 0$,
where $m$ is $m_1$ or $m_2$, for deriving formulas for the  ensemble
averaged $M_3$ and $M_4$. The final formulas are given in Appendix B and
details of the derivations will be reported elsewhere \cite{Ma-11a}. The
following results are inferred from the results in Appendix B.

It is seen from Eq. (\ref{eq.pty9a}),  $\gamma_1(m_1,m_2)$ will be non-zero
only when $\alpha \neq 0$ and the $\tau$ dependence is weak. Also, it is
seen that for $N_+ = N_-$, $\gamma_1(m_1,m_2) = - \gamma_1(m_2,m_1)$.
Similarly,  Eq. (\ref{eq.pty9b}) shows that for $N_+ = N_-$,
$\gamma_2(m_1,m_2) = \gamma_2(m_2,m_1)$. In the dilute limit, with some
approximations as discussed after Eq. (\ref{eq.pty9b}), the expression for
$\gamma_2(m_2,m_1)$ is given by  Eq. (\ref{eq.pty9c}). This shows that, for
$\alpha << \tau$ or $\tau << \alpha$, the $C_1$ and $C_2$ in Eq.
(\ref{eq.pty9c}) will be negligible and then, $\gamma_2 \sim -4/m$ for $m_1
= m_2 = m/2$ and $N_1 = N_2 = N$.  This is same as the result for spinless
fermion EGOE(2) \cite{Mo-73,Mo-75} and shows that for a range of
$(\tau,\alpha)$ values, $\rho^{m_1,m_2}(E)$ will be close to  Gaussian.
Moreover, to the extent that Eq. (\ref{eq.pty9c}) applies,  the density
$\rho^{m_1,m_2}(E)$ is a convolution of the densities generated by $X(2)$ and
$D(2)$ operators.  Let us add that the binary correlation results presented
in Appendix B,  with further extensions, will be useful in the study of
partitioned EGOE discussed in \cite{Ko-01,Ko-99}. 

\section{Results and Discussion}

In order to proceed with the numerical calculations, we need to have some
idea of the range of the parameters $(\tau, \alpha, m/N_+, N_+/N_-)$.
Towards this end, we have used realistic nuclear effective interactions in
$sdfp$ \cite{sdfp} and $fpg_{9/2}$ \cite{fpg} spaces and calculated the
variances $v_a^2$, $v_b^2$, $v_c^2$,  $v_d^2$ for these interactions. Note
that it is easy to identify  the matrices $A$, $B$, $C$ and $D$ given the
interaction matrix elements $\lan (j_1 j_2) J T \mid V \mid (j_3 j_4)
JT\ran$. To calculate the mean-squared matrix elements $v^2$'s, we put the
diagonal two-particle matrix elements to be zero and use the weight factor
$(2J+1)(2T+1)$.  Assuming that $\Delta=3$ MeV and $5$ MeV (these are
reasonable values for $A=20-80$ nuclei), we obtain $\tau \sim 0.09-0.24$ and
$\alpha \sim (0.9-1.3) \times \tau$. These deduced values of $\alpha$ and
$\tau$ clearly point out that one has to go beyond the highly restricted
ensemble employed in \cite{Pa-08} and it is necessary to consider the more
general EGOE(1+2)-$\pi$ defined in Section II.  Similarly, for $sdfp$ and
$fpg_{9/2}$ spaces  $N_+/N_- \sim 0.5-2.0$. Finally, for nuclei with $m$
number of valence nucleons (particles or holes) where  $sdfp$ or $fpg_{9/2}$
spaces are appropriate, usually $m \lazz N_+$ or $N_-$, whichever is lower.
Given these, we have selected the following examples: $(N_+,N_-,m) =
(8,8,4)$, $(8,8,5)$, $(10,6,4)$, $(10,6,5)$, $(6,10,4)$, $(6,10,5)$,
$(8,8,6)$, $(6,6,6)$, $(7,7,7)$ and $(7,7,6)$. To go beyond the matrix
dimensions $\sim 5000$ with $100$ members is not feasible at present with
the HPC cluster that is used for all the calculations.  Most of the
discussion in this paper is restricted to $N = N_+ + N_- = 16$ and $m << N$
as in this dilute limit it is possible to  understand the ensemble results
better. Following the nuclear examples mentioned above, we have chosen
$\tau=0.05,\;0.1,\;0.2,\;0.3$ and $\alpha/\tau = 0.5, \;1.0, \;1.5$. We will
make some comments on the results for other $(\tau,\alpha)$ values at
appropriate places. 

Now we will present the results for (i) the form of the  $+$ve and $-$ve
parity state densities $I_+(E)$ and $I_-(E)$ respectively, (ii) the parity
ratios $I_-(E)/I_+(E)$ vs $E$ where $E$ is the excitation energy of the
system  and (iii) the probability for $+$ve parity ground states generated
by the EGOE(1+2)-$\pi$ ensemble. 

\subsection{Gaussian form for fixed-$\pi$ state densities}

\begin{figure}
    \centering
    \subfigure
    {
        \includegraphics[width=4in,height=5.5in,angle=-90]{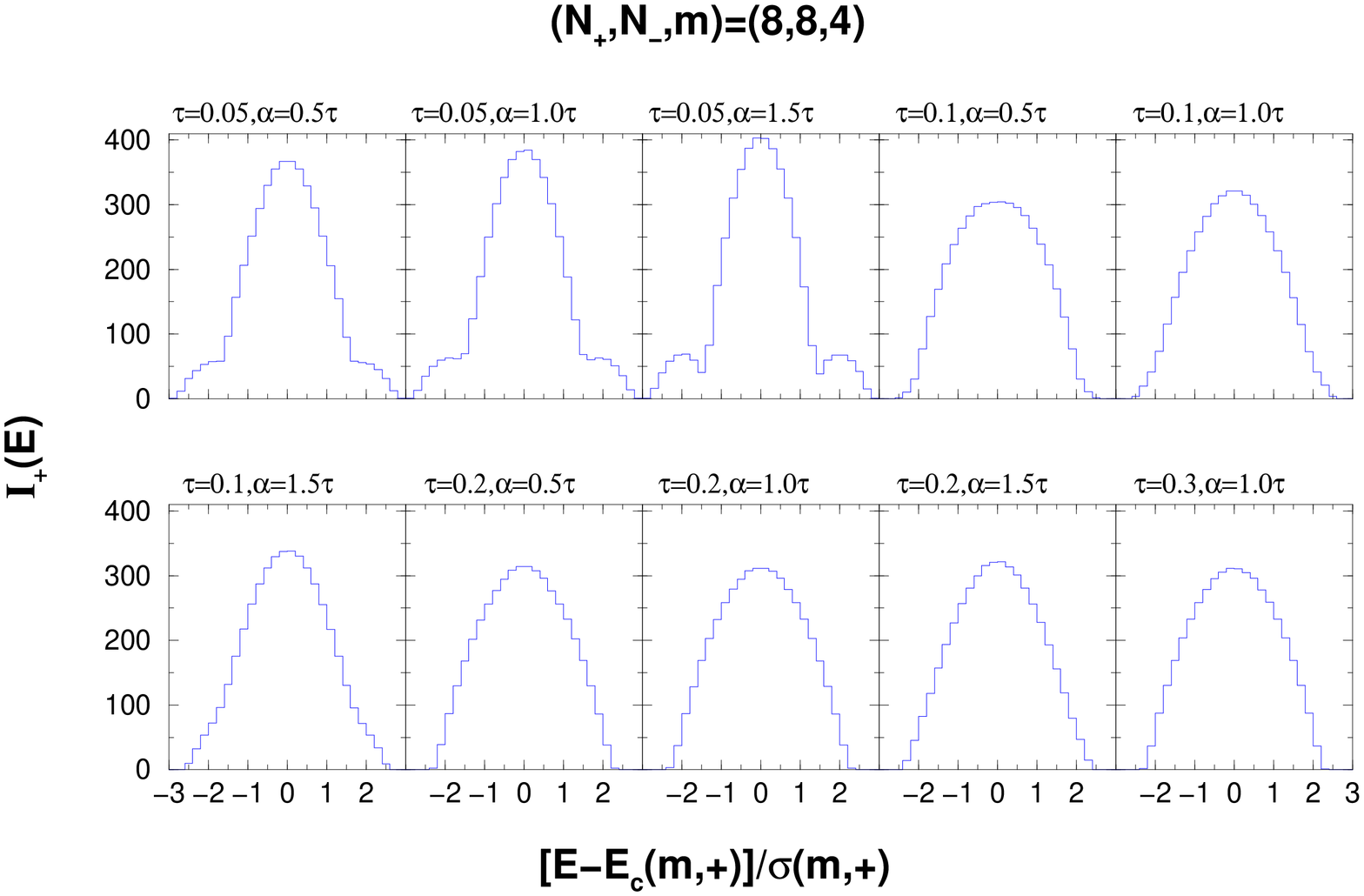}
        \label{den884-1}
    }
    \\
    \subfigure
    {
        \includegraphics[width=4in,height=5.5in,angle=-90]{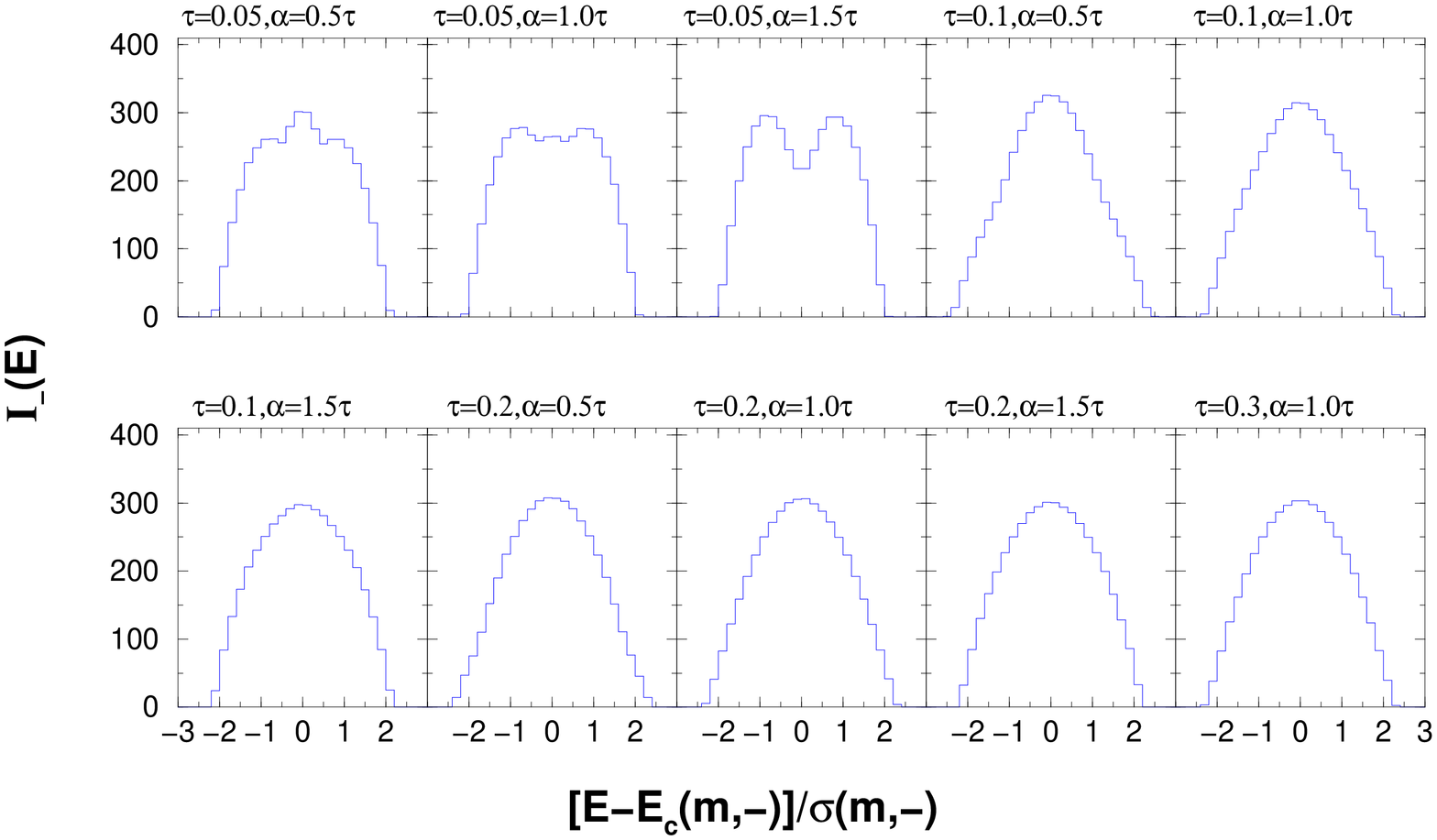}
        \label{den884-2}
    }
    \caption{(Color online) Positive and negative parity state densities for
    various $(\tau,\alpha)$ values for $(N_+,N_-,m) = (8,8,4)$ system. See
    text for details.}
    \label{den884}
\end{figure}

\begin{figure}
    \centering
    \subfigure
    {
        \includegraphics[width=4.5in,height=3.75in]{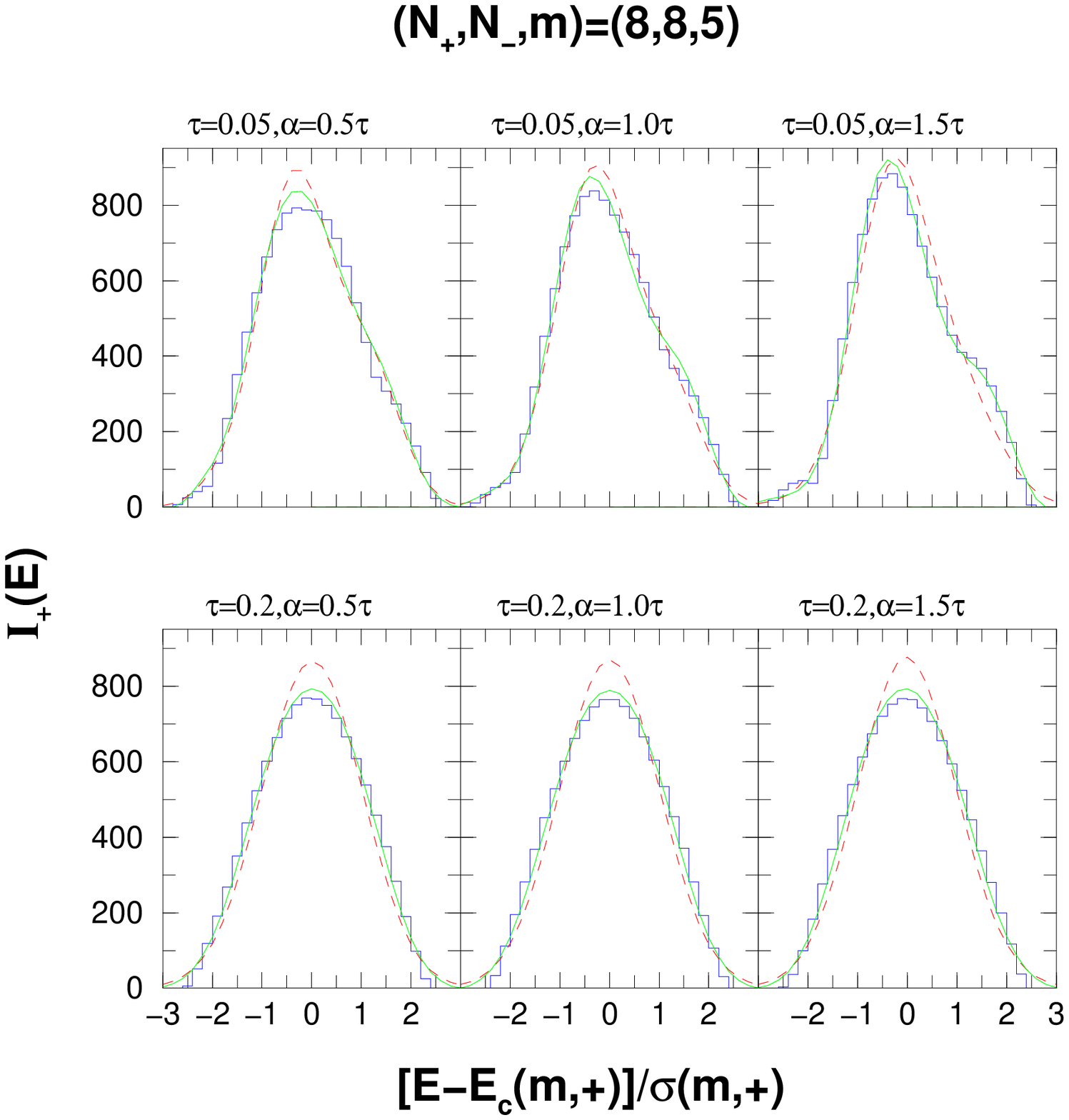}
        \label{den885-1}
    }
    \\
    \subfigure
    {
        \includegraphics[width=4.5in,height=3.75in]{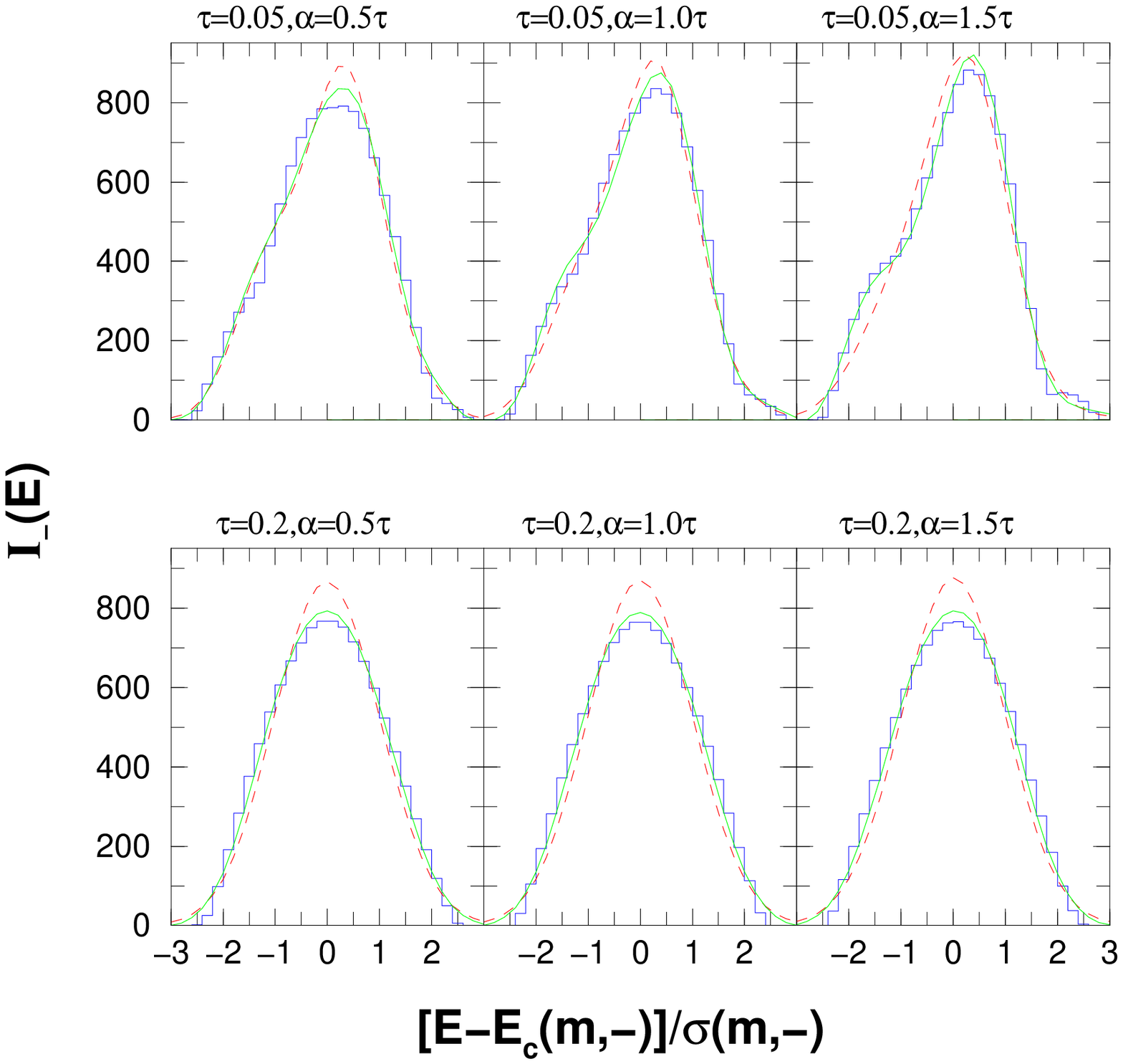}
        \label{den885-2}
    }
    \caption{(Color online) Positive and negative parity state densities 
    for various $(\tau,\alpha)$ values for $(N_+,N_-,m) = (8,8,5)$ 
    system.  Histograms are numerical ensemble results. The
    dashed (red) curve corresponds to Gaussian form for $\rho^{m_1,m_2}(E)$ in
    Eq. (\ref{eq.densty}) and similarly, solid (green) curve corresponds to 
    Edgeworth corrected Gaussian form with $\gamma_1(m_1,m_2)$ and
    $\gamma_2(m_1,m_2)$ obtained using the results in Appendix B. 
    See text for details.}
    \label{den885}
\end{figure}

\begin{figure}
    \centering
    \subfigure
    {
        \includegraphics[width=5.4in,height=3.75in]{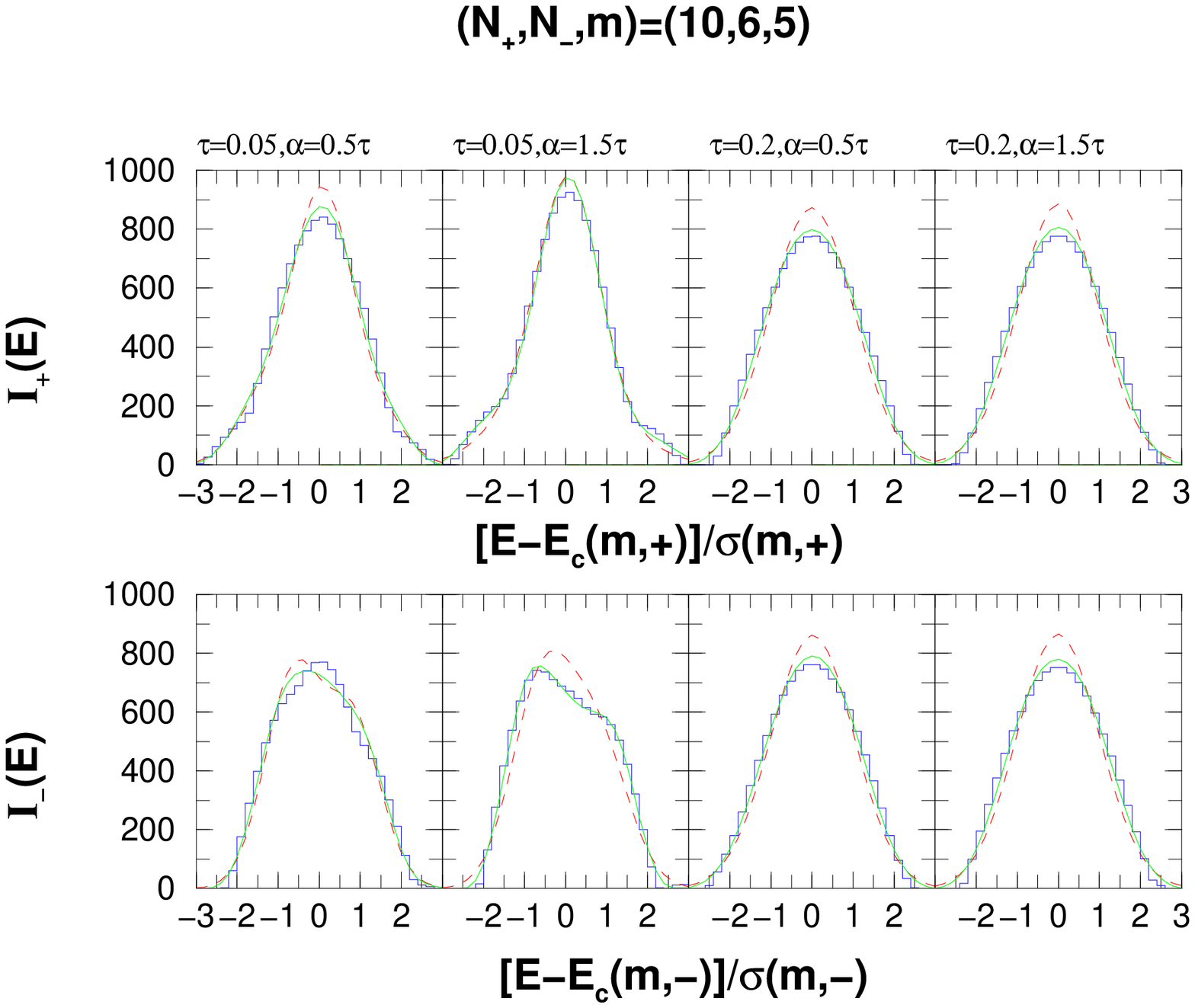}
        \label{den1065-1}
    }
    \\
    \subfigure
    {
        \includegraphics[width=5.4in,height=3.75in]{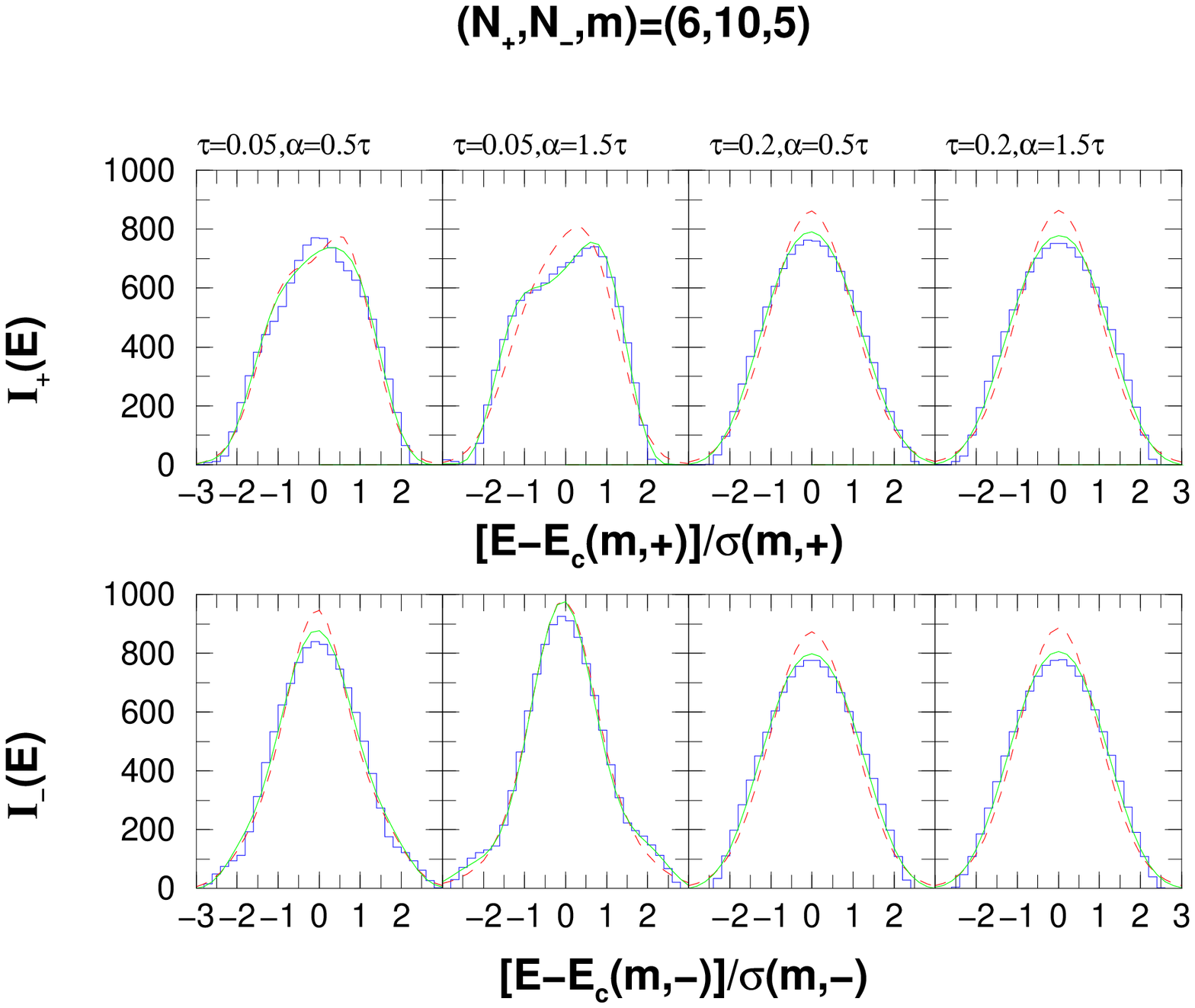}
        \label{den1065-2}
    }
    \caption{(Color online) Positive and negative parity state densities 
    for various $(\tau,\alpha)$ values for $(N_+,N_-,m) = (10,6,5)$ 
    and $(6,10,5)$ systems. Histograms are numerical ensemble results. The
    dashed (red) curve corresponds to Gaussian form for $\rho^{m_1,m_2}(E)$ in
    Eq. (\ref{eq.densty}) and similarly, solid (green) curve corresponds to 
    Edgeworth corrected Gaussian form with $\gamma_1(m_1,m_2)$ and
    $\gamma_2(m_1,m_2)$ obtained using the results in Appendix B. 
    See text for details.}
    \label{den1065}
\end{figure}

Using the method discussed in Sec. II, we have numerically constructed in
$+$ve and $-$ve parity spaces EGOE(1+2)-$\pi$ ensembles of random matrices
consisting of 100 Hamiltonian matrices in large number of examples, i.e. for
$(N_+,N_-,m)$ and $(\tau,\alpha)$ parameters mentioned above. Diagonalizing
these matrices, ensemble averaged eigenvalue (state) densities, 
\be
\overline{I_\pm(E)} = \overline{\lan\lan\delta(H-E)\ran\ran^\pm} \;,
\label{eq.denpty2}
\ee
are constructed. From now on, we drop the `overline' symbol when there is no
confusion.  Results are shown for $(N_+,N_-,m)=(8,8,4)$, $(8,8,5)$,
$(10,6,5)$ and $(6,10,5)$ for several values of $(\tau,\alpha)$ in Figs.
\ref{den884}, \ref{den885} and \ref{den1065}. To construct the fixed-parity 
eigenvalue densities, we first make the centroids $E_c(m,\pm)$ of  all the
members of the ensemble to be zero and variances $\sigma^2(m,\pm)$ to be
unity, i.e. for each member we have the  standardized  eigenvalues $\we =
[E-E_c(m,\pm)]/\sigma(m,\pm)$. Then, combining all the $\we$ and using a
bin-size $\Delta \we=0.2$, histograms for  $I_\pm(E)$ are generated. It is
seen that the state densities are multimodal for small $\tau$ values and for
$\tau \geq 0.1$, they are unimodal and close to a Gaussian.  Note that in
our examples, $\alpha=(0.5-1.5) \times \tau$.

\begin{figure}
\includegraphics[width=5in,height=5in]{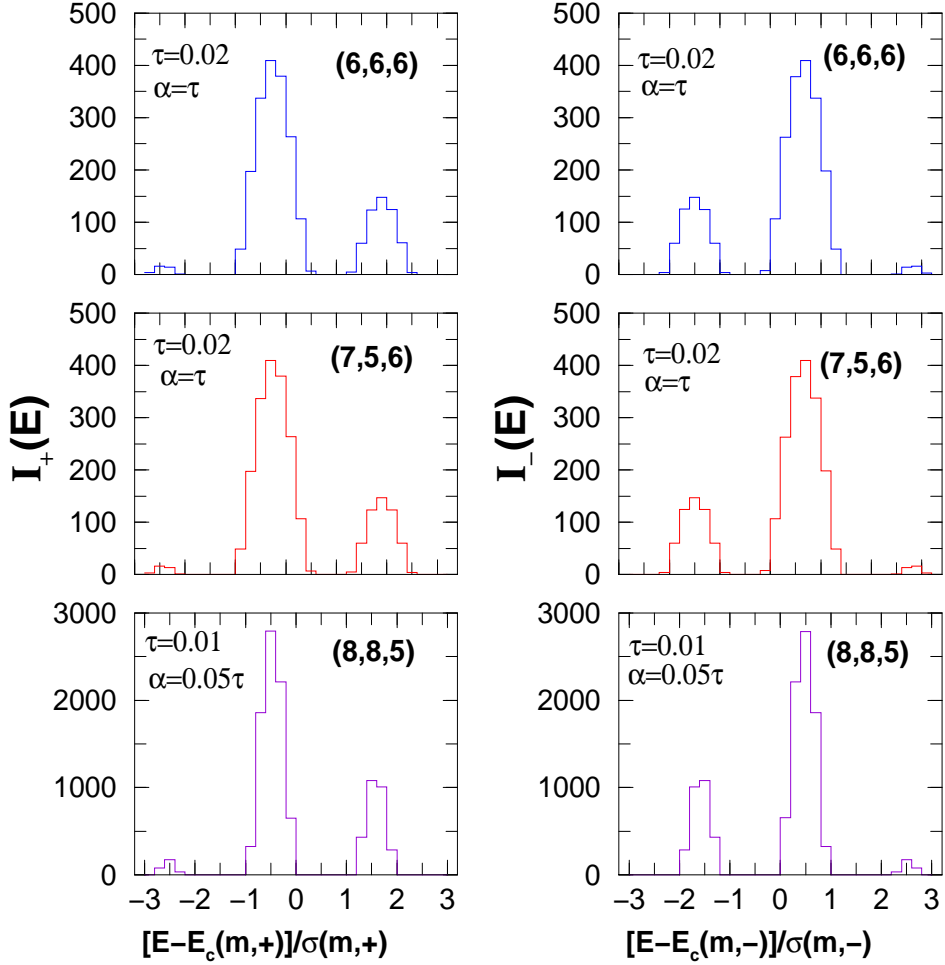}
\caption{(Color online) Positive and negative parity state densities for 
some small values of $(\tau,\alpha)$. The $(N_+,N_-,m)$ values are given 
in the figures. See text for details.}
\label{smtau}
\end{figure}

For $V(2)=0$, the eigenvalue densities will be a sum of spikes  at
$0,\;2\Delta,\;4\Delta,\;\ldots$ for $+$ve parity densities and similarly at
$\Delta,\;3\Delta,\;5\Delta,\;\ldots$ for $-$ve parity densities. As we
switch on $V(2)$, the spikes will spread due to the matrices $A$, $B$ and
$C$ in Fig. \ref{fig1} and mix due to the matrix $D$. The variance
$\sigma^2(m_1,m_2)$ can be written as, 
\be
\sigma^2(m_1,m_2) = \sigma^2(m_1,m_2 \to m_1,m_2) + \sigma^2(m_1,m_2 \to m_1
\pm 2,m_2 \mp 2)\;.
\label{eq.varm1m2}
\ee
The internal variance $\sigma^2(m_1,m_2 \to m_1,m_2)$ is due to $A$, $B$ and
$C$ matrices and it receives contribution only from the $\tau$ parameter.
Similarly, the external variance $\sigma^2(m_1,m_2 \to m_1 \pm 2,m_2 \mp 2)$
is due to the matrix $D$ and  it receives contribution only from the 
$\alpha$ parameter. When we switch on $V(2)$, as the ensemble averaged
centroids generated by $V(2)$ will be zero, the positions of the spikes will
be largely unaltered. However, they will start spreading and mixing as
$\tau$ and $\alpha$  increase. Therefore, the density will be multimodal
with the modes well separated for very small $(\tau,\alpha)$ values.  Some
examples for this are shown in Fig. \ref{smtau}. As $\tau$ and $\alpha$
start increasing from zero, the spikes spread and will start overlapping for
$\sigma(m_1,m_2) \gazz \Delta$. This is the situation with $\tau=0.05$ shown
in Figs. \ref{den884}, \ref{den885} and \ref{den1065}. However, as $\tau$
increases (with $\alpha \sim \tau$), the densities start becoming unimodal
as seen from the $\tau=0.1$ and $0.2$ examples. Also, the $m$ dependence is
not strong as seen from the Figs. \ref{den884}, \ref{den885} and
\ref{den1065}. Now we will discuss the comparison of the ensemble results 
with the smoothed densities constructed using $E_c(m_1,m_2)$,
$\sigma^2(m_1,m_2)$, $\gamma_1(m_1,m_2)$ and $\gamma_2(m_1,m_2)$. 

\begin{figure}
    \centering
    \subfigure
    {
        \includegraphics[width=5.5in,height=3.5in]{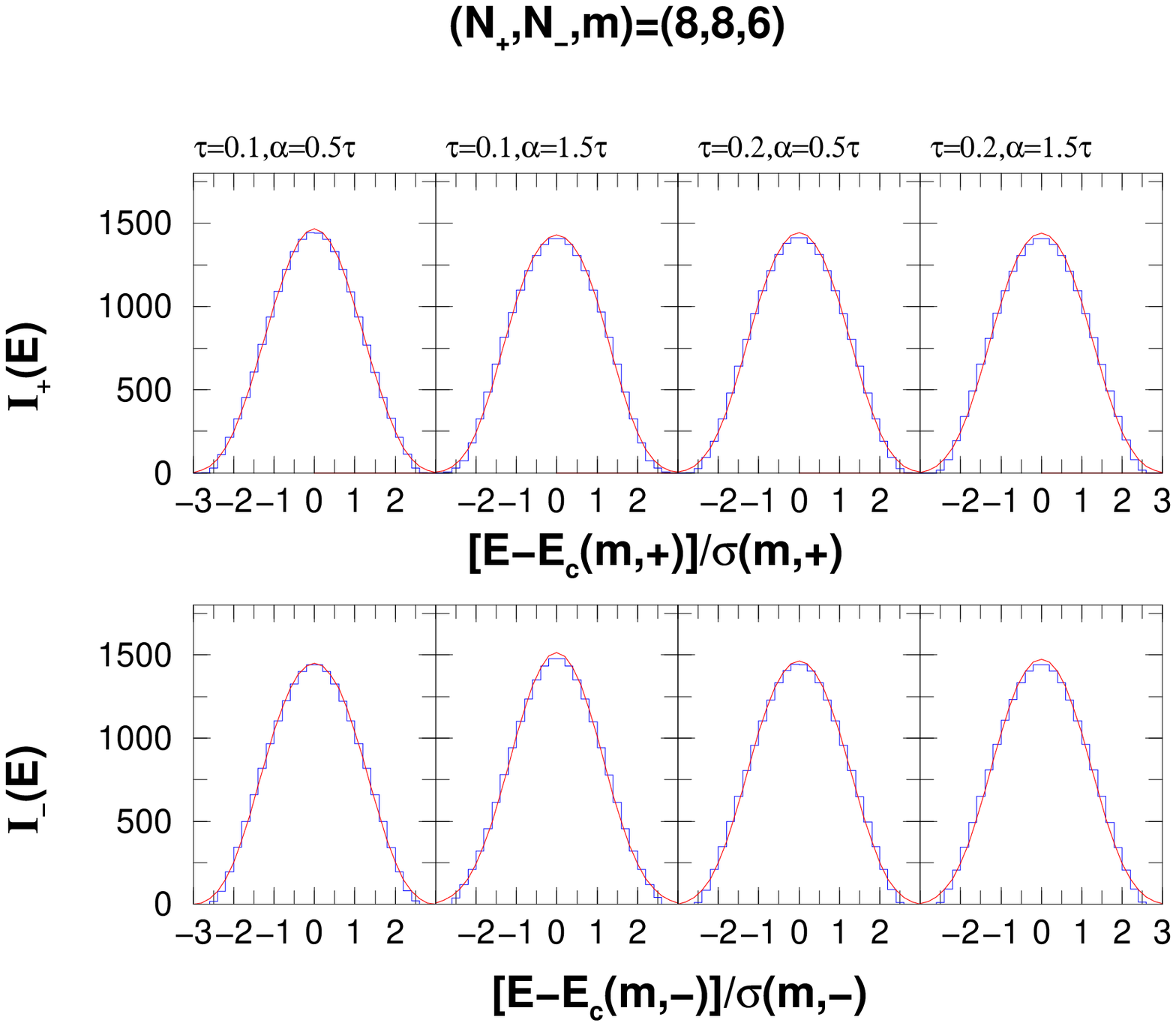}
        \label{den886-1}
    }
    \\
    \subfigure
    {
        \includegraphics[width=5.5in,height=3.5in]{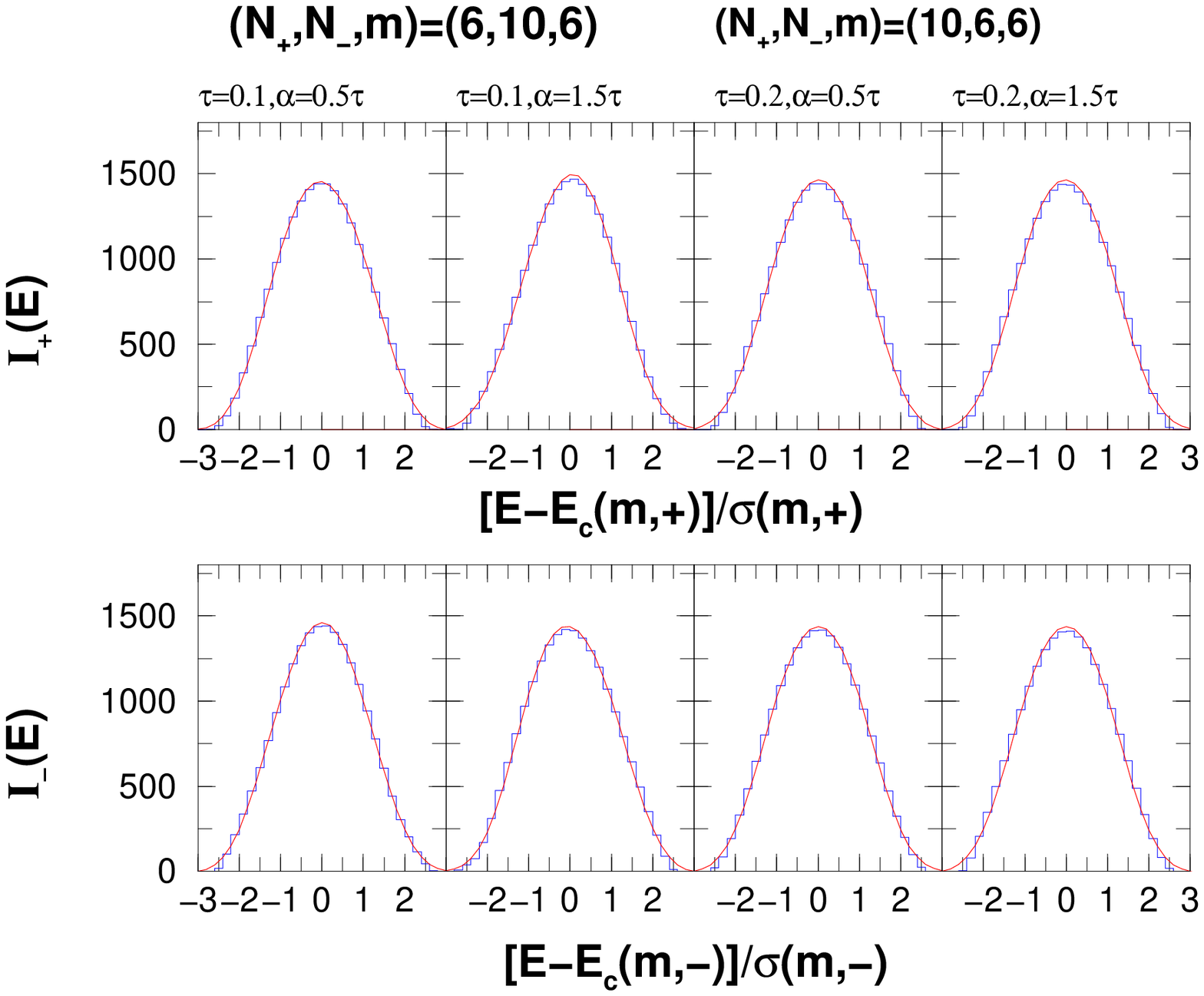}
        \label{den886-2}
    }
    \caption{(Color online) Positive and negative parity state densities for
    various $(\tau,\alpha)$ values for $(N_+,N_-,m)=(8,8,6)$, $(6,10,6)$
    and $(10,6,6)$ systems. Smoothed curves (solid red lines) are obtained 
    using fixed-$(m_1,m_2)$ partial densities. See text for details.}
    \label{den886}
\end{figure}

As the particle numbers in the examples shown in Figs. \ref{den884},
\ref{den885} and \ref{den1065} are small, the excess parameter
$\gamma_2(m,\pi) \sim -0.7$ to $-0.8$ (skewness parameter $\gamma_1(m,\pi)
\sim 0$ in all our examples). Therefore the densities are not very close to
a Gaussian form. It has been well established that the  ensemble averaged
eigenvalue density takes Gaussian form in the case of spinless fermion (as
well as boson) systems and also for  the embedded ensembles extending to
those with good quantum numbers; see \cite{Br-81,Ko-01,Go-10,Ma-10}  and
references therein.  Thus, it can be anticipated that Gaussian form is 
generic for the state densities or more appropriately,  for the partial
densities $\rho^{m_1,m_2}(E)$ generated by EGOE(1+2)-$\pi$ for some range
of $(\tau,\alpha)$ values. Results for the fixed-$\pi$ densities for
$(N_+,N_-,m)=(8,8,6)$, $(6,10,6)$ and $(10,6,6)$ systems are shown in Fig.
\ref{den886}.  The smoothed $+$ve and $-$ve parity densities are a sum of
the partial  densities $\rho^{m_1,m_2}(E)$,
\be
\rho_\pm(E) = \dis\frac{1}{d_\pm} \dis\sum^\pr_{m_1,m_2} 
d(m_1,m_2) \rho^{m_1,m_2}(E)\;. 
\label{eq.parden}
\ee
Note that the summation in Eq. (\ref{eq.parden}) is over $m_2$ even for
$+$ve parity density and similarly over $m_2$ odd for $-$ve parity density.
Here $\rho_\pm(E)$ as well as $\rho^{m_1,m_2}(E)$ are normalized to unity.
However, in practice, the densities normalized to dimensions are needed and
they are denoted, as used earlier, by $I_\pm(E)$ and $I^{m_1,m_2}(E)$ 
respectively,
\be
I_\pm(E) = d_\pm \rho_\pm(E) = \dis\sum^\pr_{m_1,m_2} 
I^{m_1,m_2}(E) \;;\;\;\;\; I^{m_1,m_2}(E) = d(m_1,m_2) \rho^{m_1,m_2}(E)\;.
\label{eq.densty}
\ee
We employ the Edgeworth (ED) form that includes  $\gamma_1$ and $\gamma_2$
corrections to the Gaussian partial densities $\rho_{\cg}^{m_1,m_2}(E)$. 
Then 
$$
\rho^{m_1,m_2}(E) \to \rho_{\cg}^{m_1,m_2}(E) \to \rho_{ED}^{m_1,m_2}(E)
$$ 
and in terms of the standardized variable $\we$, the ED form is given by
\be
\barr{rcl}
\eta_{ED}(\we) & = &
\eta_{\cg}(\we)
\l\{1+\l[{\dis\frac{\gamma_1}{6}}He_3\l(\we\r)\r]+
\l[{\dis\frac{\gamma_2}{24}}He_4\l(\we\r) +
{\dis\frac{\gamma_1^2}{72}} He_6\l(\we\r) \r]\r\}\;;\\ \\
\eta_{\cg}(\we) & = &
\dis\frac{1}{\sqrt{2\pi}}\exp
\l(-\dis\frac{\we^2}{2}\r) \;.
\earr \label{eq.gau1}
\ee
Here, $He$ are Hermite polynomials: $He_3(x) = x^3 - 3x$, $He_4(x) = x^4 -
6x^2 + 3$ and $He_6(x) = x^6 - 15x^4 + 45x^2 - 15$.  Using Eqs.
(\ref{eq.parden})  and (\ref{eq.gau1}) with exact centroids and variances
given by the propagation formulas in Section III and  the binary correlation
results for $\gamma_1$ and $\gamma_2$ as given by the formulas in Appendix
B,   the smoothed $+$ve and $-$ve parity state densities are constructed. 
We put $\eta_{ED}(\we)=0$ when  $\eta_{ED}(\we)<0$. It is clearly seen from
Fig. \ref{den886} that the sum of partial densities, with the partial
densities represented by ED corrected Gaussians, describe extremely  well
the exact fixed-$\pi$ densities in these examples.  Therefore, for the
$(\tau,\alpha)$ values in the range determined by nuclear $sdfp$ and
$fpg_{9/2}$ interactions, i.e. $\tau \sim 0.1-0.3$ and $\alpha \sim 0.5 \tau
- 2 \tau$,  the partial densities can be well represented by ED corrected
Gaussians and total densities are also close to ED corrected Gaussians.
Unlike Fig. \ref{den886}, densities in Figs. \ref{den884}, \ref{den885} and
\ref{den1065} show, in many cases, strong departures form Gaussian form.
Therefore, it is important to test how well Eq. (\ref{eq.densty}) with ED
corrected Gaussian for $\rho^{m_1,m_2}(E)$ describes the numerical results
for $I_\pm(E)$. We show this comparison for all the densities in Figs.
\ref{den885} and \ref{den1065}. It is clearly seen that the agreements  with
ED corrected Gaussians are good in all the cases. Therefore, the large
deviations from the Gaussian form for $I_\pm(E)$ arise mainly because of the
distribution of the centroids [this involves dimensions of the $(m_1,m_2)$
configurations] of the partial densities involved. It is possible that the
agreements in Figs. \ref{den885} and \ref{den1065} may become more perfect
if we employ, for the partial densities, some non-canonical forms defined by
the first four moments as given for example in \cite{Gr-95,Jo-06b}. However,
as these forms are not derived using any random matrix ensemble, we haven't
used these for the partial densities in our present investigation. In
conclusion, for the physically relevant range of $(\tau,\alpha)$ values, the
propagation formulas for centroids and variances given by Eqs.
(\ref{eq.cent}) and (\ref{eq.var}) or alternatively with $E_c(m_1,m_2) =
m_2\Delta$ and Eq. (\ref{eq.eavar})  along with the EGOE(1+2)-$\pi$ ensemble
averaged $\gamma_1(m_1,m_2)$ and $\gamma_2(m_1,m_2)$ formulas (obtained
using the binary correlation approximation as given in Appendix B) can be
used to construct  fixed-$\pi$ state densities for larger $(N_+,N_-,m)$
systems. 

\subsection{Parity ratios for state densities}

\begin{figure}
\includegraphics[width=4in,height=5in]{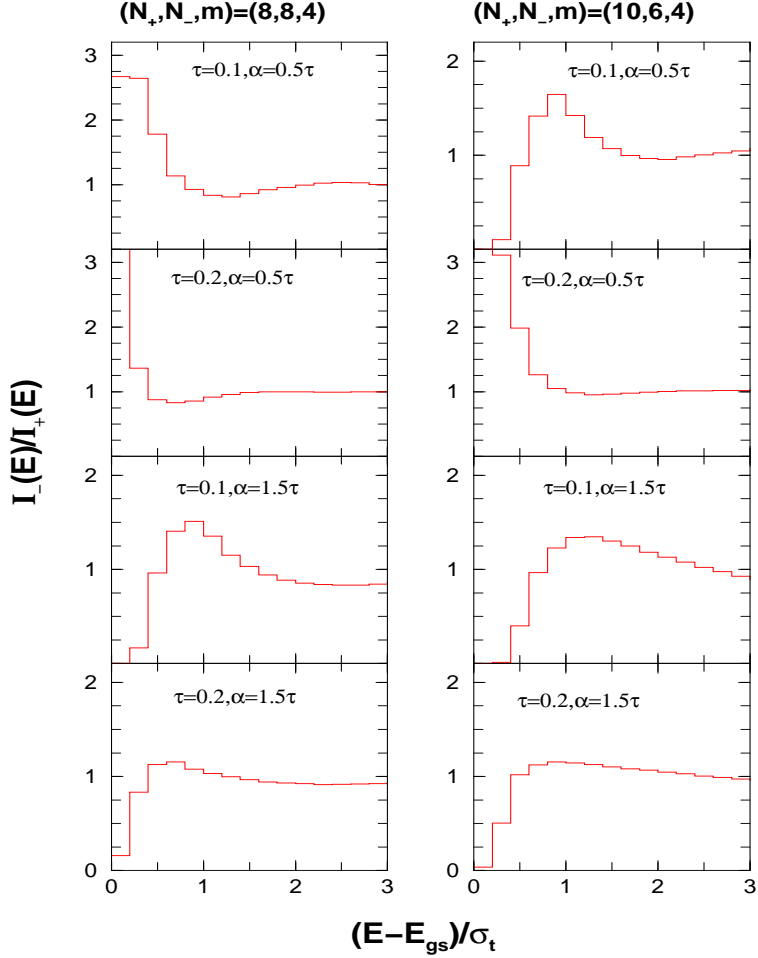}
\caption{(Color online) Parity ratios for various $(\tau,\alpha)$ values for
$(N_+,N_-,m) = (8,8,4)$ and $(10,6,4)$ systems. See text for details.}
\label{pr884}
\end{figure}

\begin{figure}
    \centering
    \subfigure
    {
        \includegraphics[width=5.5in,height=4in]{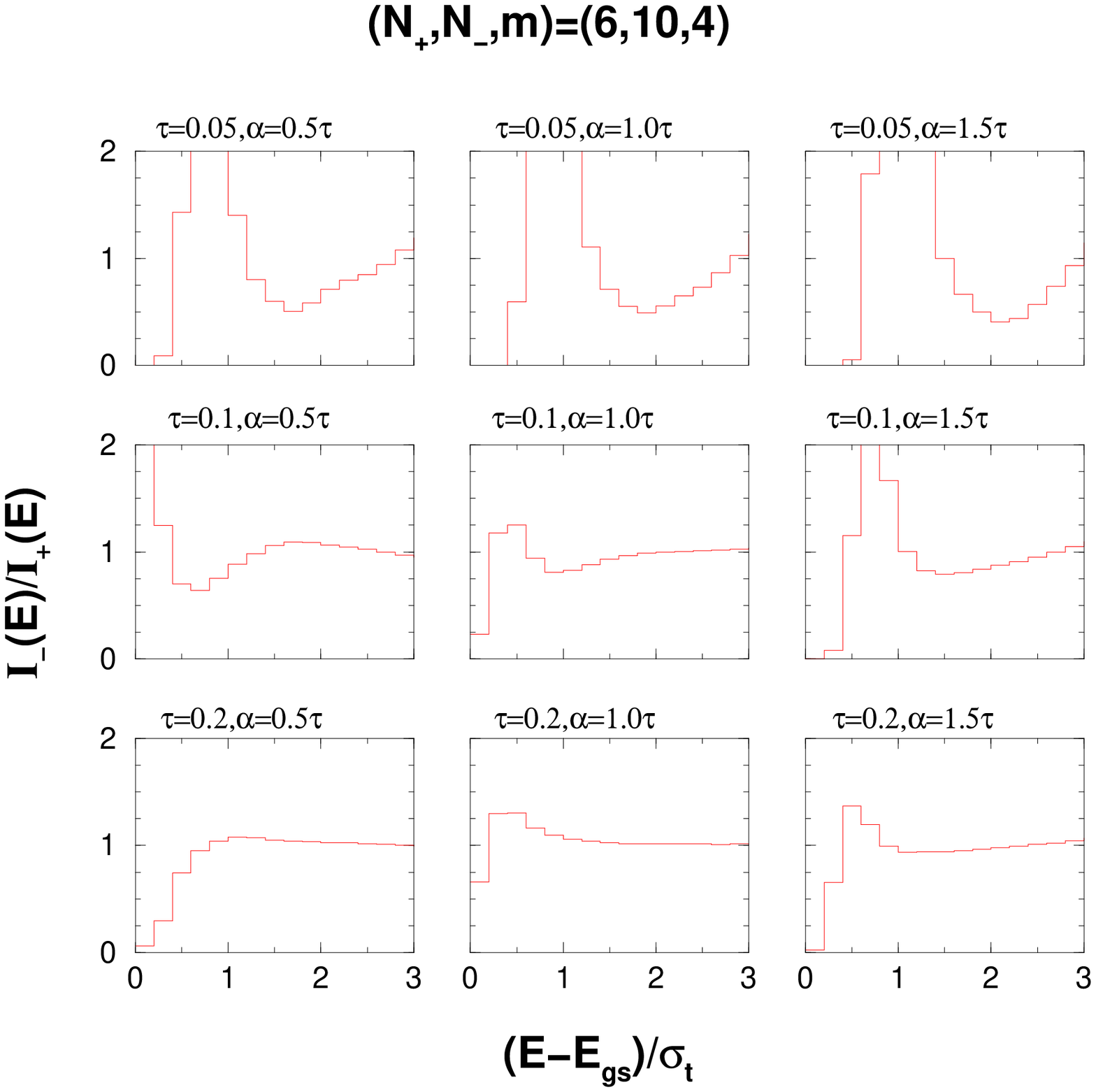}
        \label{pr610-1}
    }
    \\
    \subfigure
    {
        \includegraphics[width=5.5in,height=4in]{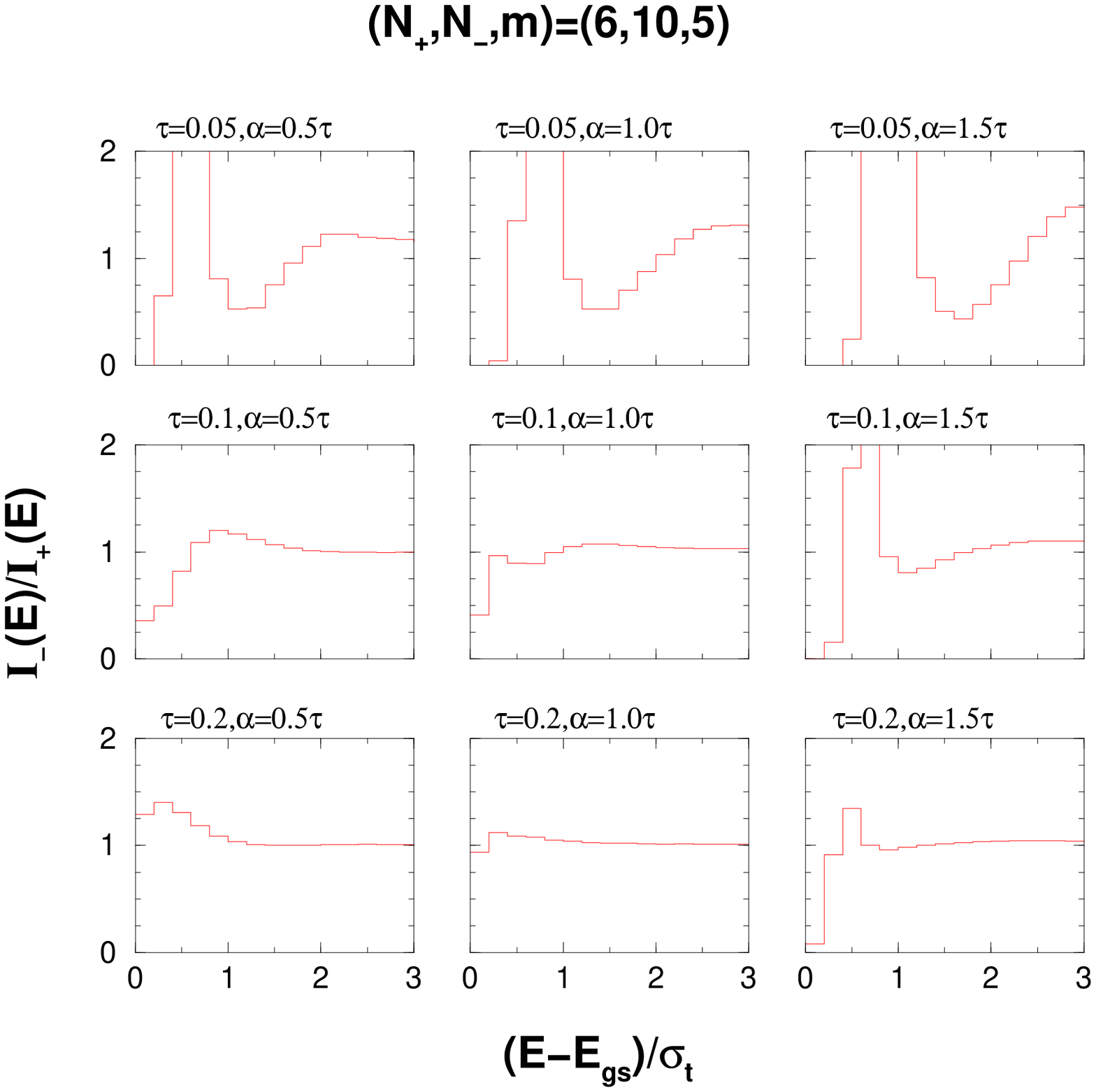}
        \label{pr610-2}
    }
    \caption{(Color online) Parity ratios for various $(\tau,\alpha)$ values
    for $(N_+,N_-,m) = (6,10,4)$ and $(6,10,5)$  systems. See text for
    details.}
    \label{pr6105}
\end{figure}

\begin{figure}
\includegraphics[width=5in,height=5.5in]{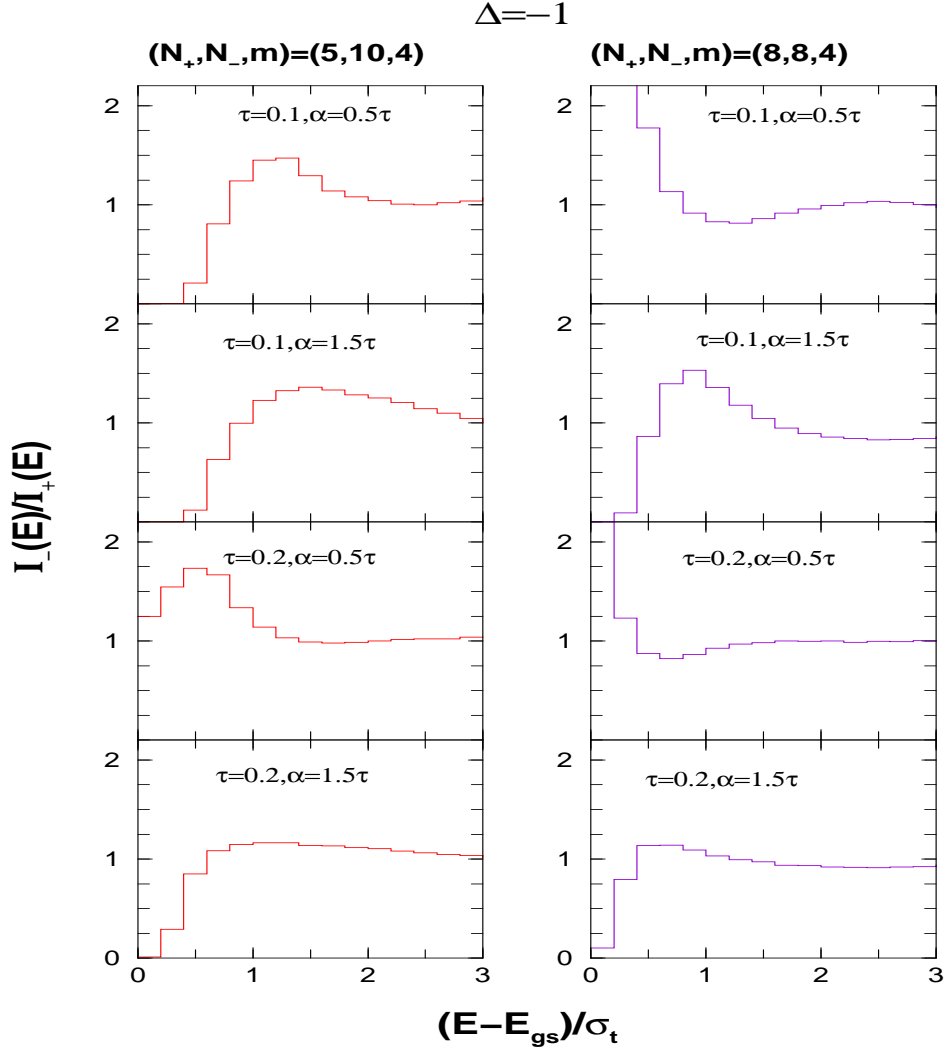}
\caption{(Color online) Parity ratios for some values of $(\tau,\alpha)$ 
with $\Delta=-1$ for $(N_+,N_-,m) = (5,10,4)$ and $(8,8,4)$ systems. 
See text for details.}
\label{del-1}
\end{figure}

\begin{figure}
\includegraphics[width=5in,height=5in]{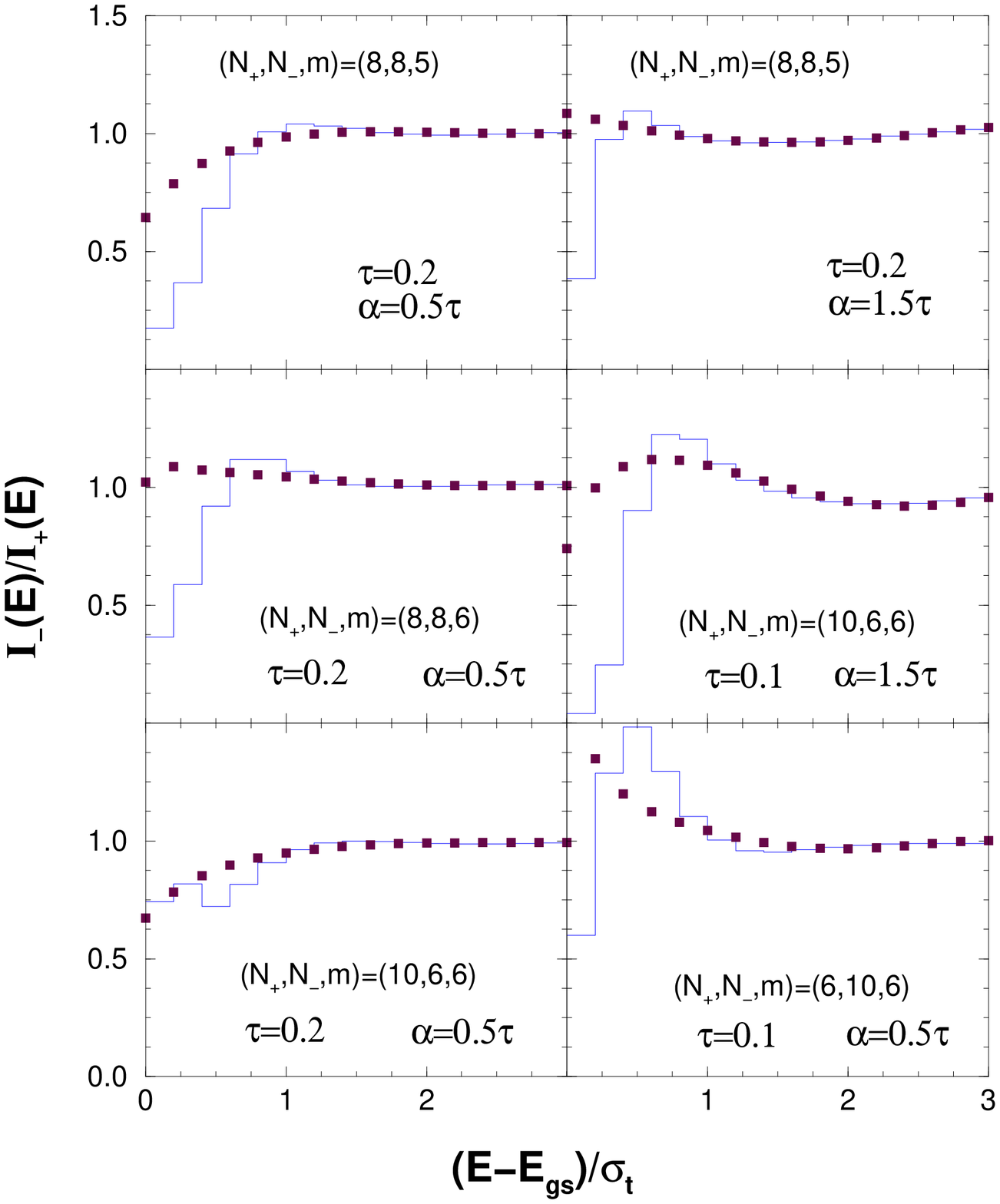}
\caption{(Color online) Parity ratios for various $(\tau,\alpha)$ values and
for various $(N_+,N_-,m)$ systems. Filled squares (brown)
are obtained using fixed-$(m_1,m_2)$ partial densities. See text for details.}
\label{pr886}
\end{figure}

As stated in the beginning, parity ratio of state densities at a given
excitation energy $(E)$ is a quantity of considerable interest in nuclear
structure. For the systems shown in Figs. \ref{den884}, \ref{den885} and
\ref{den1065} and also for many other systems,  we have studied the parity
ratios and the results are shown in Figs. \ref{pr884}-\ref{pr886}. As the
parity ratios need to be  calculated at a given value of excitation energy
$E$, we measure the eigenvalues in both $+$ve and $-$ve parity spaces with
respect to the absolute ground state energy $E_{gs}$ of the $N = N_+ + N_-$
system.  Thus, $E_{gs}$ is defined by taking all the $+$ve and $-$ve parity
eigenvalues of all members of the ensemble and choosing the lowest of all
these.  The ground state energy can also be determined by averaging the
$+$ve and $-$ve parity ground state energies over the ensemble and then the
ground state energy is minimum of the two. It is seen that the results for
parity ratios are essentially independent of the choice of $E_{gs}$ and thus
we employ absolute ground state energy in our calculations. We use the
ensemble averaged total ($+$ve and $-$ve eigenvalues combined) spectrum
width $\sigma_t$ of the system for scaling. The total widths $\sigma_t$ can
be calculated also  by using $E_c(m_1,m_2)$ and $\sigma^2(m_1,m_2)$.
Examples for $\sigma_t$ are shown in Table \ref{widths} and they are in good
agreement with the results obtained using the simple formula given by 
Eq. (\ref{eq.eavar}). We use the variable $\bee = (E-E_{gs})/\sigma_t$ for
calculating parity ratios. Starting with $E_{gs}$ and using a bin-size of
$\Delta\bee = 0.2$, we have calculated the number of states $I_+(\bee)$ with
$+$ve parity and also the number of states $I_-(\bee)$  with $-$ve parity 
in a given bin and then the ratio $I_-(\bee)/I_+(\bee)$ is the parity ratio.
Note that the results in  Figs. \ref{pr884}-\ref{pr886} are shown for
$\bee=0-3$ as the spectrum span is $\sim 5.5\sigma_t$. To go beyond the
middle of the spectrum,  for real nuclei, one has to include more sp levels
(also a finer splitting of the $+$ve and $-$ve parity levels may be needed)
and therefore, $N_+$ and $N_-$ change. Continuing with this, one obtains the
Bethe form for nuclear level densities \cite{KH-10}.

General observations  from Figs. \ref{pr884}-\ref{pr886} are as follows. (i)
The parity ratio $I_-(\bee)/I_+(\bee)$ will be zero up to an energy
$\bee_0$. (ii) Then, it starts increasing and becomes larger than unity at
an  energy $\bee_m$. (iii) From here on, the parity ratio decreases and
saturates quickly to unity from an energy $\bee_1$. In these examples, 
$\bee_0 \lazz 0.4$, $\bee_m \sim 1$ and $\bee_1 \sim 1.5$.  It is seen that
the curves shift towards left as $\tau$ increases. Also the position of the
peak shifts to much larger value of $\bee_m$ and equilibration gets delayed
as $\alpha$ increases for a fixed $\tau$ value.  Therefore for larger
$\tau$, the energies $(\bee_0, \bee_m, \bee_1)$ are smaller compared to
those for a smaller $\tau$. The three transition energies also depend on
$(N_+,N_-,m)$. We have also verified, as shown in Fig. \ref{del-1}, that the
general structure of the parity ratios will remain same even when we change
$\Delta \to -\Delta$ (i.e. $-$ve parity sp states below the $+$ve parity sp
states). For $(N_+,N_-,m)=(8,8,4)$ system, results for $\Delta = 1$ are
given in Fig. \ref{pr884} and they are almost same as the results with
$\Delta=-1$ given in Fig. \ref{del-1}. The general structures (i)-(iii) are
clearly seen in the numerical examples shown in \cite{Mo-07} where a method
based on the Fermi-gas model has been employed. If $\sigma_t \sim 6-8$ MeV,
equilibration in parities is expected around $E \sim 8-10$ MeV and this is
clearly seen in the examples in \cite{Mo-07}. It is also seen from Fig.
\ref{pr6105} that the equilibration is quite poor for very small values of
$\tau$ and therefore comparing with the results in \cite{Mo-07}, it can be
argued that very small values of $\tau$ are ruled out for nuclei. Hence, it
is plausible to conclude that generic results for parity ratios can be
derived using EGOE(1+2)-$\pi$ with reasonably large $(\tau,\alpha)$ values.
Let us add that the interpretations in \cite{Mo-07} are based on the
occupancies of the sp orbits while in the present work, they are in terms of
$\tau$ and $\alpha$ parameters.

Using the smoothed $I^\pm(\bee)$, constructed as discussed in Section IV A,
smoothed forms for parity ratios are calculated as follows. Starting with
the absolute ground state energy $E_{gs}$ and using a bin-size of
$\Delta\bee = 0.2$, $+$ve and $-$ve parity  densities in a given energy bin 
are obtained and their ratio is the parity ratio at a given $\bee$. We have
chosen the examples where $I_+$ and $I_-$ are close to Gaussians. It is seen
from Fig. \ref{pr886} that the agreement with exact results is good for
$\bee \gazz 0.5$. However, for smaller $\bee$, to obtain a good agreement
one should have a better prescription for determining the tail part of the
$\rho^{m_1,m_2}(\bee)$ distributions. Developing the theory for this is
beyond the scope of the present paper as this requires more complete 
analytical treatment of the ensemble.  

\subsection{Probability for $+$ve parity ground states}

Papenbrock and Weidenm\"{u}ller used the $\tau \to \infty$, $\alpha = \tau$
limit of EGOE(1+2)-$\pi$ for several $(N_+,N_-,m)$ systems to study the
probability $(R_+)$  for $+$ve parity ground states over the ensemble
\cite{Pa-08}. As stated before, this exercise was motivated by shell model
results with random interaction giving preponderance of  $+$ve parity ground
states \cite{Zh-04a}. The numerical calculations in \cite{Pa-08} showed
considerable variation ($18-84$\%) in $R_+$. In addition, they gave a
plausible proof that in the dilute limit $[m << (N_+,N_-)]$, $R_+$ will
approach $50$\%. Combining these, they argued that the observed
preponderance of $+$ve parity ground states could be a finite size (finite
$N_+$, $N_-$, $m$) effect. For the extended EGOE(1+2)-$\pi$ considered in
the present work, where the $\tau \to \infty$ and $\alpha = \tau$
restriction is relaxed, as we will discuss now, $R_+$ can reach $100$\%.

\begin{figure}
\includegraphics[width=5in,height=6.5in]{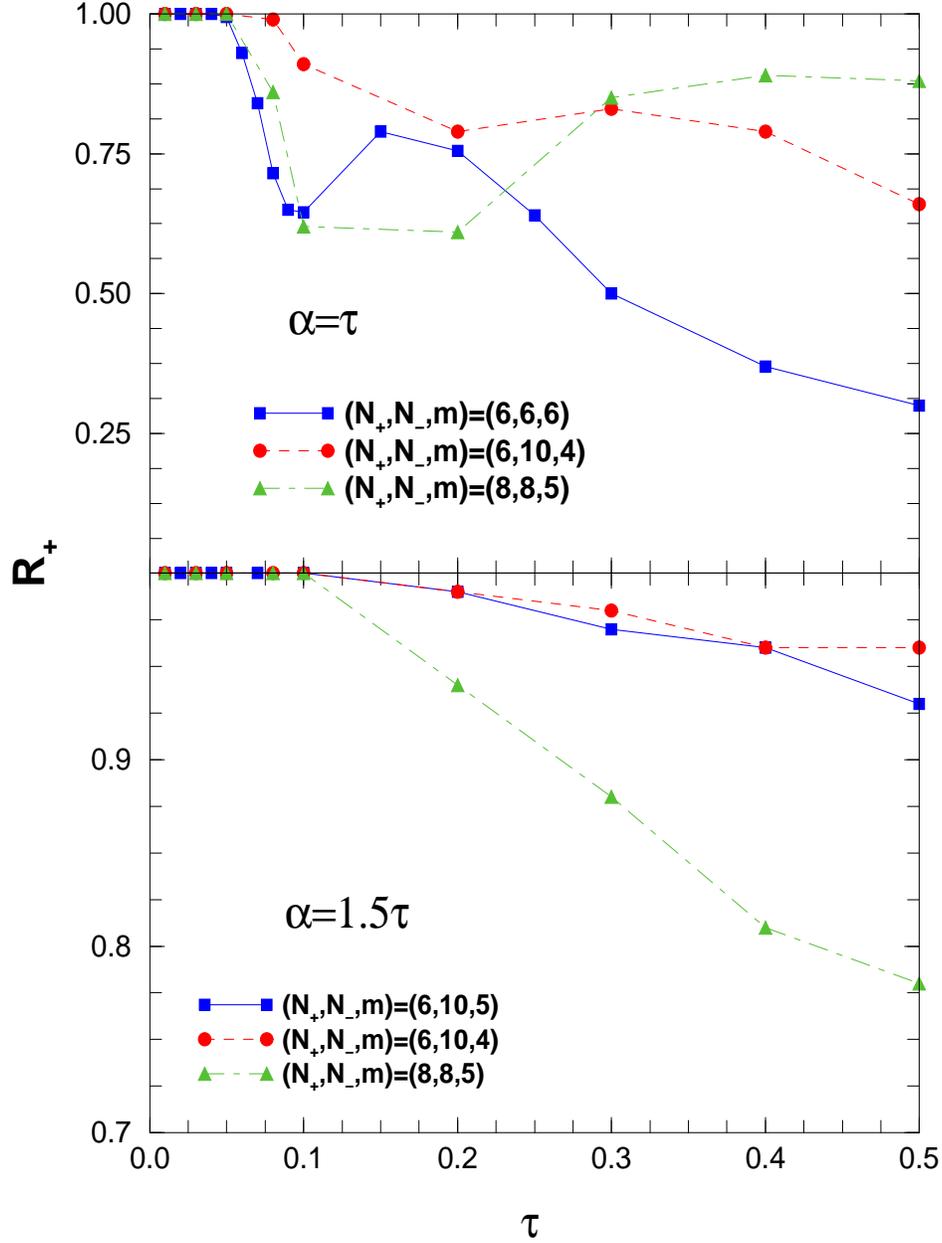}
\caption{(Color online) Probability $(R_+)$ for $+$ve parity ground states 
for various $(\tau,\alpha)$ values and for various $(N_+,N_-,m)$ 
systems. See text for details.}
\label{rplus}
\end{figure}

For EGOE(1+2)-$\pi$ with $\tau \sim 0$, clearly one will get  $R_+ = 100$\%
(for even $m$ and $m << N_+,N_-$) and therefore it is of interest to study
$R_+$ variation with $(\tau,\alpha)$. We have carried out calculations using
a 200 member ensemble for $(N_+,N_-,m)=(6,6,6)$ and 100 member ensembles for
$(8,8,5)$, $(6,6,6)$,  $(6,10,4)$ and $(6,10,5)$ systems. In these
calculations, we use $\alpha=\tau$ and $1.5\tau$. The results are shown in
Fig. \ref{rplus}.  For $\alpha=\tau$, the results are as follows. For $\tau
\lazz 0.04$, we have $R_+ \sim 100$\% and then $R_+$ starts decreasing with
some fluctuations between $\tau=0.1$ and $0.2$. The origin of these
fluctuations is not clear. As $\tau > 1$ is not realistic, we have
restricted the $R_+$ calculations to $\tau \leq 1$. We see from the figure
that EGOE(1+2)-$\pi$ generates $R_+ \gazz 50$\% for $\tau \leq 0.3$
independent of $(N_+,N_-,m)$. Also, $R_+$ decreases much faster with $\tau$
and reaches $\sim 30$\% for $\tau=0.5$ for $(N_+,N_-,m)=(6,6,6)$. For $m <
(N_+,N_-)$, the decrease in $R_+$ is slower. If we increase $\alpha$, from
the structure of the two-particle $H$ matrix in Fig. \ref{fig1}, we can
easily infer that the width of the lowest $+$ve parity $(m_1,m_2)$ unitary
configuration becomes much larger compared to the lowest $-$ve parity
unitary configuration (see  Table \ref{widths} for examples). Therefore,
with increasing $\alpha$ we expect $R_+$ to increase and this is clearly
seen in Fig. \ref{rplus}. Thus $\alpha \gazz \tau$ is required for $R_+$ to
be large. A quantitative description of $R_+$ requires the construction of
$+$ve and $-$ve parity state densities more accurately in the tail region
and the  theory for this is not yet available. 

\section{Conclusions and future outlook}

In the present work, we have introduced a generalized EGOE(1+2)-$\pi$ 
ensemble for identical fermions and its construction follows from EGOE(1+2)
for spinless fermion systems. Using this generalized EE, we have not only
studied $R_+$, as it was done by Papenbrock and Weidenm\"{u}ller
\cite{Pa-08} using a simpler two-body ensemble with parity, but also studied
the form of fixed-$\pi$ state densities and parity ratios which are
important nuclear structure quantities. Numerical examples (see Figs.
\ref{den884}-\ref{den1065} and \ref{den886}), with the range of the various
parameters in the model fixed using realistic nuclear effective
interactions, are used to show that the fixed-$\pi$ state densities in
finite dimensional spaces are of Gaussian form for sufficiently large values
of the mixing parameters $(\tau,\alpha)$. The random matrix model also
captures the essential features of parity ratios as seen in the method based
on non-interacting Fermi-gas model reported in \cite{Mo-07}. We also found
preponderance of $+$ve parity ground states for $\tau \lazz 0.5$ and $\alpha
\sim 1.5 \tau$. In addition, for constructing fixed-$\pi$ Gaussian densities
we have derived an easy to understand propagation formula [see Eq.
(\ref{eq.var})] for the spectral variances  of the partial densities
$\rho^{m_1,m_2}(E)$ that generate $I_+$ and $I_-$. Similarly, for calculating
the corrections to the Gaussian forms, formulas for skewness $\gamma_1$ and
excess $\gamma_2$ of the partial densities $\rho^{m_1,m_2}(E)$ are derived
using the binary correlation approximation (see Appendix B  for the
formulas). The smoothed densities constructed using Edgeworth corrected
Gaussians are shown to  describe the numerical results for $I_\pm(E)$ [for
$(\tau,\alpha)$ values in the range defined by nuclear $sdfp$ and
$fpg_{9/2}$ interactions - see beginning of Sec. IV] and also the parity
ratios at energies away from the ground state. Numerical results presented
for parity ratios at lower energies show that a better theory for the tails
of the partial densities is needed (see Figs. \ref{pr884}-\ref{pr886}).
Thus, the results in the present paper represent considerable progress in
analyzing EGOE(1+2)-$\pi$ ensemble going much beyond the analysis presented
in \cite{Pa-08}.

Results in the present work are largely numerical and they clearly show that
developing a complete analytical theory, going beyond the results presented
in Sections III and Appendix B, for EGOE(1+2)-$\pi$ is important. In future,
it is important to investigate EGOE(1+2)-$\pi$ for proton-neutron systems
and then we will have four unitary orbits (two for protons and two for
neutrons). In addition, by including non-degenerate $+$ve and $-$ve parity
sp states the model could be applied to nuclei for predicting parity ratios.
This extended EGOE(1+2)-$\pi$ model with protons and neutrons occupying
different sp states will be generated by a 10$\times$10 block matrix for
$V(2)$ in two-particle spaces. Therefore, parametrization of this ensemble
is more complex. Analysis of this extended EGOE(1+2)-$\pi$ is for future.  

\acknowledgments

All the calculations in the present paper are carried out using the HPC
cluster resource at Physical Research Laboratory. Thanks are due to E.R.
Prahlad and G. Vignesh for collaboration in the initial stages. Thanks are
also due to S. Tomsovic for some useful discussions and to the referee for
some useful comments.

\renewcommand{\theequation}{A\arabic{equation}}
\setcounter{equation}{0}   
\section*{APPENDIX A} 

Let us consider a system of $m$ fermions in $N$ sp states with a (1+2)-body
Hamiltonian $H=h(1)+V(2)$ where $h(1)=\sum_i \epsilon_i\, {\hat{n}}_i$  and
$V(2)$ is defined by the two-body matrix elements   $V_{ijkl}= \lan kl
\mid V(2) \mid ij \ran$.  With respect to the $U(N)$ group, the two-body
interaction $V(2)$ can be separated into scalar ($\nu=0$), effective
one-body ($\nu=1$) and irreducible two-body ($\nu=2$) parts 
\cite{CFT,Wo-86,Ko-01,KH-10},
\be
\barr{l}
V^{\nu=0} = \dis\frac{\hat{n}(\hat{n} -1)}{2}
{\overline{V}}\;\;;\;\;\;\; {\overline{V}}=
\dis\binom{N}{2}^{-1} \; \dis\sum_{i<j} V_{ijij} \;,\\
V^{\nu=1} = \dis\frac{{\hat{n}}-1}{N-2} \dis\sum_{i,j} \zeta_{i,j}
a^\dagger_ia_j\;\;;\;\;\zeta_{i,j}=
\l[\dis\sum_{k} V_{kikj}\r] -\l[(N)^{-1}\;\dis\sum_{r,s} 
V_{rsrs}\r] \delta_{i,j} \;,\\
V^{\nu=2} = V - V^{\nu=0} - V^{\nu=1}\;\; \Longleftrightarrow \;\;
V^{\nu=2}_{ijkl}\;\;; \\
V^{\nu=2}_{ijij} = V_{ijij}- \overline{V} - (N-2)^{-1} \l(\zeta_{i,i}+
\zeta_{j,j} \r)\;\;,\\ 
V^{\nu=2}_{ijik} = V_{ijik} - (N-2)^{-1} \zeta_{j,k}\;\;\;\mbox{for}\;\; 
j \neq k \;,\\
V^{\nu=2}_{ijkl} = V_{ijkl} \;\;\;\mbox{for all other cases} \;.
\earr \label{eq.a1}
\ee
Similar to Eq. (\ref{eq.a1}), the $h(1)$ operator will have $\nu=0,1$ parts,
\be
\barr{l}
h^{\nu=0}= {\overline{\epsilon}}\;{\hat{n}}\;,\;\;\;
{\overline{\epsilon}}=(N)^{-1} \dis\sum_i \epsilon_i \;,\\
h^{\nu=1} =
\dis\sum_i\,\epsilon^1_i {\hat{n}}_i\;\;,\;\;\;\epsilon^1_i =
\epsilon_i-{\overline{\epsilon}} \;.
\earr \label{eq.a2}
\ee
Then the propagation equations for the $m$-particle centroids and 
variances are \cite{CFT,Wo-86,Ko-01,KH-10},
\be
\barr{rcl}
E_c(m) & = & \lan H \ran^m =  m\,\overline{\epsilon} + 
\dis\binom{m}{2}\;\overline{V} \;,\\ \\
\sigma^2(m) & = & \lan H^2 \ran^m - \l[ E_c(m) \r]^2 \\ \\
& = & 
\dis\frac{m(N-m)}{N(N-1)} \;\;\dis\sum_{i,j}\;
\l\{\epsilon^1_i \delta_{i,j} + \dis\frac{m-1}{N-2} \zeta_{i,j}\r\}^2 \\ \\
& + & \dis\frac{m(m-1)(N-m)(N-m-1)}{N(N-1)(N-2)(N-3)} \;\lan\lan 
\l(V^{\nu=2} \r)^2 \ran\ran^2 \;.
\earr \label{eq.a3}
\ee

\begin{table}[htp]
\caption{Exact results for skewness and excess parameters for fixed-$\pi$
eigenvalue densities $I_\pm(E)$ compared with the binary correlation
results (in the table, called `Approx'). For exact results, we have used the
eigenvalues obtained from  EGOE(1+2)-$\pi$ ensembles with 100 members. The
binary correlation results are obtained using Eqs.
(\ref{eq.pty1})-(\ref{eq.pty9}) and extension of  Eq. (\ref{eq.cenvar}). See
text for details.}
\begin{tabular}{ccrrrrrrrr}
\hline
 &  & \multicolumn{4}{c}{$\gamma_1(m,\pi)$} & 
\multicolumn{4}{c}{$\gamma_2(m,\pi)$} \\ \cline{3-10}
$(N_+,N_-,m)$ & $(\tau,\alpha/\tau)$ & 
\multicolumn{2}{c}{Exact} & \multicolumn{2}{c}{Approx} &
\multicolumn{2}{c}{Exact} & \multicolumn{2}{c}{Approx} \\ \cline{3-10}
 & & $\pi=+\;$ & $\pi=-\;$ & $\pi=+\;$ & $\pi=-\;$ 
 & $\pi=+\;$ & $\pi=-\;$ & $\pi=+\;$ &  $\pi=-$ \\
\hline
$(8,8,5)$ 
 & $(0.05,0.5)$ & $0.15$ & $-0.15$ & $0.15$ & $-0.15$ & $-0.52$ & $-0.52$ &
 $-0.52$ & $-0.52$ \\
 & $(0.05,1.0)$ & $0.16$ & $-0.16$ & $0.16$ & $-0.16$ & $-0.50$ & $-0.50$ &
 $-0.50$ & $-0.50$ \\
 & $(0.05,1.5)$ & $0.18$ & $-0.17$ & $0.18$ & $-0.18$ & $-0.46$ & $-0.46$ &
 $-0.46$ & $-0.46$ \\
 & $(0.2,0.5)$  & $-0.03$ & $0.03$ & $-0.03$ & $0.03$ & $-0.71$ & $-0.71$ &
 $-0.71$ & $-0.71$ \\
 & $(0.2,1.0)$  & $-0.01$ & $0.01$ & $-0.01$ & $0.01$ & $-0.73$ & $-0.73$ &
 $-0.74$ & $-0.74$ \\
 & $(0.2,1.5)$  & $0.02$ & $-0.02$ & $0.02$ & $-0.02$ & $-0.72$ & $-0.72$ &
 $-0.73$ & $-0.73$ \\
$(10,6,5)$ 
 & $(0.05,0.5)$ & $-0.06$ & $0.09$ & $-0.07$ & $0.09$ & $-0.26$ & $-0.76$ &
 $-0.26$ & $-0.75$ \\
 & $(0.05,1.5)$ & $-0.04$ & $0.15$ & $-0.05$ & $0.15$ & $-0.01$ & $-0.86$ &
 $-0.01$ & $-0.86$ \\
 & $(0.2,0.5)$  & $0.01$ & $-0.04$ & $0.01$ & $-0.04$ & $-0.73$ & $-0.69$ &
 $-0.73$ & $-0.69$ \\
 & $(0.2,1.5)$  & $0.01$ & $0.02$ & $0.01$ & $0.02$ & $-0.69$ & $-0.75$ &
 $-0.70$ & $-0.75$ \\
$(6,10,5)$ 
 & $(0.05,0.5)$ & $-0.09$ & $0.07$ & $-0.09$ & $0.07$ & $-0.76$ & $-0.26$ &
 $-0.75$ & $-0.26$ \\
 & $(0.05,1.5)$ & $-0.15$ & $0.05$ & $-0.15$ & $0.05$ & $-0.86$ & $-0.01$ &
 $-0.86$ & $-0.01$ \\
 & $(0.2,0.5)$  & $0.04$ & $-0.01$ & $0.04$ & $-0.01$ & $-0.68$ & $-0.73$ &
 $-0.69$ & $-0.73$ \\
 & $(0.2,1.5)$  & $-0.02$ & $-0.01$ & $-0.02$ & $-0.01$ & $-0.75$ & $-0.69$ &
 $-0.75$ & $-0.70$ \\
\hline
\end{tabular}
\label{c5t1}
\end{table}

\renewcommand{\theequation}{B\arabic{equation}}
\setcounter{equation}{0}   
\section*{APPENDIX B} 

For the EGOE(1+2)-$\pi$ Hamiltonian defined in Eq. (\ref{eq.ham}), we have
$H = h(1) + V(2) = h(1) + X(2) + D(2)$ with  $X(2) = A \oplus B \oplus C$ is
the direct sum of the spreading matrices $A$, $B$ and $C$ and $D(2) = D +
\wD$ is the off-diagonal mixing matrix as shown in Fig. \ref{fig1}. Here,
$\wD$ is the transpose of the matrix $D$. With the sp energies defining the
mean field $h(1)$ as given in Eq. (\ref{eq.ham}), the first moment $M_1$  of
$\rho^{m_1,m_2}(E)$ is trivially,
\be
M_1(m_1,m_2) = \overline{\lan (h + V) \ran^{m_1,m_2}} = m_2 \;,
\label{eq.pty1}
\ee
as $\lan h^r \ran^{m_1,m_2} = (m_2)^r$ and  $\overline{\lan V
\ran^{m_1,m_2}} = 0$. By extending the binary correlation method
to traces over two-orbit configurations, we have derived formulas
for the second, third and fourth order traces giving $M_r(m_1,m_2)$,
$r=2-4$. It is important to mention that the presence of the mixing matrix
$D$ makes the derivations lengthy. Therefore,  we give only the final
formulas in this paper and discuss elsewhere the details of the derivations
\cite{Ma-11a}. The second moment $M_2$ is,
\be
\barr{rcl}
M_2(m_1,m_2) & = & 
\overline{\lan (h + V)^2 \ran^{m_1,m_2}} = \lan h^2 \ran^{m_1,m_2} +
\overline{\lan V^2 \ran^{m_1,m_2}} \\
& = & (m_2)^2+ \cx(m_1,m_2) + \cd(m_1,m_2) + \bd(m_1,m_2)\;; \\ \\
\cx(m_1,m_2) = \overline{\lan X^2 \ran^{m_1,m_2}} & = & \tau^2 \; 
\dis\sum_{i+j=2}
\dis\binom{\wmp+i}{i} \dis\binom{m_1}{i} \dis\binom{\wmn+j}{j}
\dis\binom{m_2}{j}\;, \\ \\
\cd(m_1,m_2)= \overline{\lan D\wD \ran^{m_1,m_2}} & = & \alpha^2 \; 
\dis\binom{m_1}{2}
\dis\binom{\wmn}{2}\;,\\ \\
\bd(m_1,m_2)=\overline{\lan \wD D \ran^{m_1,m_2}} & = & \alpha^2 \; 
\dis\binom{\wmp}{2}
\dis\binom{m_2}{2}\;.
\earr \label{eq.pty2}
\ee
Here, for brevity we have defined $\cx(m_1,m_2) =
\overline{\lan X^2 \ran^{m_1,m_2}}$, $\cd(m_1,m_2) = \overline{\lan D \wD
\ran^{m_1,m_2}}$ and  $\bd(m_1,m_2) = \overline{\lan \wD D \ran^{m_1,m_2}}$.
Note that, Eq. (\ref{eq.pty2}) gives the binary correlation formula for 
$\overline{\sigma^2(m_1,m_2)}$ and it reduces to Eq. (\ref{eq.eavar}) as
expected. Similarly, the third moment $M_3$ is
\be
\barr{rcl}
M_3(m_1,m_2) & = & \overline{\lan (h + V)^3 \ran^{m_1,m_2}} = 
\lan h^3 \ran^{m_1,m_2}
+ 2 \lan h \ran^{m_1,m_2} \overline{\lan V^2 \ran^{m_1,m_2}} \\ \\
& + & 
\overline{\lan X h X \ran^{m_1,m_2}} +
\overline{\lan D h \wD \ran^{m_1,m_2}} +
\overline{\lan \wD h D \ran^{m_1,m_2}} \\ \\
& = & (m_2)^3 + 3(m_2)\;\cx(m_1, m_2) +
(3m_2 + 2) \cd(m_1, m_2) + (3m_2 - 2)\bd(m_1, m_2) 
\earr \label{eq.pty3}
\ee
The formula for the fourth moment $M_4$ is,
\be
\barr{l}
M_4(m_1,m_2) = \overline{\lan (h + V)^4 \ran^{m_1,m_2}} =  
\lan h^4 \ran^{m_1,m_2}
+ 3 \lan h^2 \ran^{m_1,m_2} \overline{\lan V^2 \ran^{m_1,m_2}} \\ \\
+ \lan h^2 \ran^{m_1,m_2} \overline{\lan X^2 \ran^{m_1,m_2}} 
+ \overline{\lan D h^2 \wD \ran^{m_1,m_2}} +
\overline{\lan \wD h^2 D \ran^{m_1,m_2}} +
2 \;\overline{\lan h X h X \ran^{m_1,m_2}} \\ \\
+2 \;\overline{\lan h D h \wD \ran^{m_1,m_2}} +
2 \;\overline{\lan h \wD h D \ran^{m_1,m_2}} +
\overline{\lan V^4 \ran^{m_1,m_2}} \\ \\
= (m_2)^4 + 6(m_2)^2 \cx(m_1, m_2) + \l[6(m_2)^2+8(m_2)+4\r]\cd(m_1, m_2) \\ 
+ \l[6(m_2)^2-8(m_2)+4\r]\bd(m_1, m_2) + \overline{\lan V^4 \ran^{m_1,m_2}} 
\earr \label{eq.pty4}
\ee
The only unknown in Eq. (\ref{eq.pty4}) is $\overline{\lan V^4 \ran^{m_1,m_2}}$
and the expression for this is complicated,
\be
\barr{l}
\overline{\lan V^4 \ran^{m_1,m_2}} =  
\overline{\lan X^4 \ran^{m_1,m_2}} + 3 \;
\overline{\lan X^2 \ran^{m_1,m_2}} \l\{
\overline{\lan D \wD \ran^{m_1,m_2}} + \overline{\lan \wD D 
\ran^{m_1,m_2}} \r\}  +\overline{\lan D X^2 \wD \ran^{m_1,m_2}} \\ \\
+ \overline{\lan \wD X^2 D \ran^{m_1,m_2}} +
2 \; \overline{\lan X D X \wD \ran^{m_1,m_2}} + 2 \;
\overline{\lan X \wD X D
\ran^{m_1,m_2}} + \overline{\lan (D + \wD)^4 \ran^{m_1,m_2}} \\ \\
= 2 [\cx(m_1, m_2)]^2 +3 [\cx(m_1, m_2)][\cd(m_1, m_2)+\bd(m_1, m_2)] \\ \\
+ T_1(m_1,m_2) + T_2(m_1,m_2) + 2T_3(m_1,m_2) + T_4(m_1,m_2)\;.
\earr \label{eq.pty5}
\ee
where
\be
\barr{l}
T_1(m_1,m_2) =  \tau^4  \;\dis\sum_{i+j=2,\;t+u=2} F(m_1,N_1,i,t) \;
F(m_2,N_2,j,u) \;; \\ \\
F(m,N,k_1,k_2) =
\dis\sum_{s=0}^{k_2} \dis\binom{m-s}{k_2-s}^2 \; 
\dis\binom{N-m+k_1-s}{k_1} \; \dis\binom{m-s}{k_1} \; \dis\binom{N-m}{s} \;
\dis\binom{m}{s} \; \dis\binom{N+1}{s} \\ \\
\times  \dis\frac{N-2s+1}{N-s+1} \;
\dis\binom{N-s}{k_2}^{-1} \; \dis\binom{k_2}{s}^{-1} \;, \\ \\
T_2(m_1,m_2) = \cd(m_1,m_2) \cx(m_1-2,m_2+2) + \bd(m_1,m_2) \cx(m_1+2,m_2-2)
\;,\\ \\
T_3(m_1,m_2) = \tau^2 \; \alpha^2 \; \dis\sum_{i+j=2} \l[ 
\dis\binom{m_1-i}{2} \dis\binom{\wmn-j}{2} + \dis\binom{\wmp-i}{2}
\dis\binom{m_2-j}{2} \r] \\ \\ 
\times  \dis\binom{\wmp+i}{i} \dis\binom{m_1}{i} 
\dis\binom{\wmn+j}{j} \dis\binom{m_2}{j} \;, \\ \\
T_4(m_1,m_2) = \l[\cd(m_1,m_2)\r]^2 + \l[\bd(m_1,m_2)\r]^2 \\ \\
+ \cd(m_1,m_2)\l[2\;\cd(m_1-2,m_2+2) + \bd(m_1-2,m_2+2)\r] \\ \\
+ \bd(m_1,m_2)\l[2\;\bd(m_1+2,m_2-2) + \cd(m_1+2,m_2-2)\r]
+4\;\cd(m_1,m_2) \bd(m_1,m_2)\;.
\earr \label{eq.pty6}
\ee
Given the moments $M_r(m_1,m_2) = \overline{\lan H^r\ran^{m_1,m_2}}$; $r=1-4$,
the skewness and excess parameters $\gamma_1$ and $\gamma_2$ are as follows
\cite{St-87},
\be
\gamma_1(m_1,m_2) =
\dis\frac{k_3(m_1,m_2)}{[k_2(m_1,m_2)]^{3/2}}\;,\;\;\;\;\;\;
\gamma_2(m_1,m_2) = \dis\frac{k_4(m_1,m_2)}{[k_2(m_1,m_2)]^2}\;,
\label{eq.pty9ab}
\ee 
where,
\be
\barr{rcl}
k_2(m_1,m_2) & = & M_2(m_1,m_2) - M_1^2(m_1,m_2)\;, \\ 
k_3(m_1,m_2) & = & M_3(m_1,m_2) - 3\; M_2(m_1,m_2)\; M_1(m_1,m_2) + 
2\; M_1^3(m_1,m_2) \;, \\ 
k_4(m_1,m_2) & = & M_4(m_1,m_2) - 4\; M_3(m_1,m_2)\; M_1(m_1,m_2) - 
3\; M_2^2(m_1,m_2) \\ 
& + & 12\; M_2(m_1,m_2)\; M_1^2(m_1,m_2) - 6\; M_1^4(m_1,m_2) \;.
\earr \label{eq.pty9}
\ee
After carrying out the simplifications using Eqs. 
(\ref{eq.pty1})-(\ref{eq.pty9}), it is easily seen that,
\be
\gamma_1(m_1,m_2) = \dis\frac{2\l[\cd(m_1,m_2)- \bd(m_1,m_2)\r]}
{\l\{\cd(m_1,m_2) + \bd(m_1,m_2) + \cx(m_1,m_2)\r\}^{3/2}}\;.
\label{eq.pty9a}
\ee
The expression for $\gamma_2$ is more complex,
\be
\barr{l}
\gamma_2(m_1,m_2) = \l\{\bd(m_1,m_2) + \cd(m_1,m_2) + \cx(m_1,m_2) \r\}^{-2}
\\ \times
\;\l[T_1(m_1,m_2) + T_2(m_1,m_2) + 2\;T_3(m_1,m_2) + T_4(m_1,m_2) \r. \\
+ \l. \l[\bd(m_1,m_2) + \cd(m_1,m_2) \r] \l[4 -\cx(m_1,m_2) \r] - 2 \;\l[ 
\bd(m_1,m_2) + \cd(m_1,m_2) \r]^2 \r] - 1 \;.
\earr \label{eq.pty9b}
\ee
With $T_1 \sim [\cx(m_1,m_2)]^2 +  C_1(m_1,m_2)$, $T_2 = T_3 \sim
\cx(m_1,m_2)[\bd(m_1,m_2) + \cd(m_1,m_2)]$ and  $T_4 \sim 3 [\bd(m_1,m_2) +
\cd(m_1,m_2)]^2 + C_2(m_1,m_2)$ which are good in the dilute limit ($|C1|$
and $|C_2|$ will be close to zero), we have
\be
\gamma_2(m_1,m_2) = \dis\frac{C_1(m_1,m_2) + C_2(m_1,m_2) + 4\;[\bd(m_1,m_2)
+ \cd(m_1,m_2)]}
{\l\{ \bd(m_1,m_2) + \cd(m_1,m_2) + \cx(m_1,m_2) \r\}^2} \;.
\label{eq.pty9c}
\ee
Note that $C_1$ and $\cx$ depend only on $\tau$. Similarly, $C_2$ and
$(\bd,\cd)$ depend only on $\alpha$.  The $(\bd+\cd)$ term in the numerator
will contribute to $\gamma_2(m_1,m_2)$ when $\tau = 0$ and $\alpha$ is very
small. The approximation $T_2 = T_3 \sim \cx(\bd + \cd)$ is crucial in
obtaining the numerator in Eq. (\ref{eq.pty9c}) with no cross-terms
involving the  $\alpha$ and $\tau$ parameters. With this, we have $k_4$ to
be the sum of $k_4$'s coming from $X(2)$ and $D(2)$ matrices [note that, as
mentioned before,  $X(2) = A \oplus B \oplus C$ and $D(2) = D + \wD$]. 

To test the accuracy of the formulas for $M_r$ given by Eqs.
(\ref{eq.pty1})-(\ref{eq.pty6}), the binary correlation results for
$\gamma_1(m,\pm)$ and $\gamma_2(m,\pm)$  are compared with exact results
obtained using the eigenvalues from EGOE(1+2)-$\pi$ ensembles with 100
members for several values of $(N_+,N_-,m)$ and $(\tau,\alpha)$ parameters
in Table \ref{c5t1}. Extension of Eq. (\ref{eq.cenvar}) along with  the
results derived for $M_r(m_1,m_2)$ will give the binary correlation results
for $\gamma_1(m,\pm)$ and $\gamma_2(m,\pm)$. It is clearly seen from the
results in the Table that in all the examples considered, the binary
correlation results are quite close to the exact results. Similar agreements
are also seen in many other examples which are not shown in the table.

\ed